\documentclass[reprint,
nofootinbib,
 amsmath,amssymb,
 aps,
]{revtex4-2}
\usepackage{xcolor}
\usepackage{graphicx}% Include figure files
\usepackage{dcolumn}% Align table columns on decimal point
\usepackage{bm}% bold math
\begin{document}

\preprint{APS/123-QED}

%\title{Exploring the interplay between hydrodynamics and solvent quality in ring polymer solutions: A dissipative particle dynamics simulation study}
%\title{Multi-chain effects in polymer solutions with explicit hydrodynamic interactions: A dissipative particle dynamics simulation study}
\title{Mesoscopic simulations of linear and ring polymer solutions with explicit hydrodynamics under good and poor solvent conditions}
% Force line breaks with \\

% \altaffiliation[Department of Physics, Indian Institute of Technology (BHU), Varanasi, Uttar Pradesh-221005, India.}%Lines break automatically or can be forced with \\
 
\author{Ashish Kumar Singh}%
\email{assingh@sissa.it}
\affiliation{SISSA (Scuola Internazionale Superiore di Studi Avanzati), Via Bonomea 265, 34136 Trieste, Italy}%\textbackslash\textbackslash
%}%

\author{Angelo Rosa}
\email{anrosa@sissa.it}
\affiliation{SISSA (Scuola Internazionale Superiore di Studi Avanzati), Via Bonomea 265, 34136 Trieste, Italy}%\textbackslash\textbackslash

%\affiliation{
%\affiliation{
%\affiliation{%

%collaboration{CLEO Collaboration}%\noaffiliation

\date{\today}% It is always \today, today,

%%%
\begin{abstract}
%%%
We employ large-scale Dissipative Particle Dynamics simulations to investigate dilute solutions of linear polymers and unknotted, non-concatenated ring polymers in explicit solvent.
By systematically varying solvent quality, we examine the interplay between hydrodynamic interactions, chain architecture, and intermolecular association.
Under good solvent conditions, both linear and ring polymers remain expanded and well dispersed, displaying center-of-mass dynamics consistent with normal diffusion.
In poor solvents, attractive polymer-polymer interactions drive the formation of irregular aggregates characterized by partial chain collapse, substantial interpenetration, and slower dynamics.
Despite their different topologies, the two polymer architectures exhibit remarkably similar structural and dynamical responses across the solvent conditions considered.
These results indicate that solvent quality largely determines the organization and transport properties of dilute polymer solutions, whereas topological effects remain comparatively weak in the investigated regime.
%%%
\end{abstract}
%%%

%%%
\maketitle
%%%

%%%
\section{Introduction}\label{Intro}
%%%
Unknotted and non-concatenated circular (ring) polymers in bulk solution provide a well-controlled framework to investigate the effects of topology in polymer liquids~\cite{RubinsteinPRL1986,kapnistos2008unexpected,RosaPRL2014,MichielettoTurner2016,Ubertini_PRE2021,Ubertini_Macromolecules2023,TubianaPhysRep2024,Schroeder2025}.
In particular, and in comparison to linear-chain counterparts, the constraint of non-concatenation severely limits the number of accessible chain conformations once polymers begin to overlap, leading to structural and dynamical properties that can differ significantly from those of linear chains.
Beyond their fundamental interest, ring polymers are also relevant in biological systems, as several biopolymers -- including mitochondrial~\cite{Kozik2019}, plasmid~\cite{Roca2024}, and bacterial DNA~\cite{jun2006entropy,Junier2023} -- naturally occur in circular or looped forms, where topology plays a key role in cellular organization.

In dilute and semidilute solutions, polymer conformations are strongly influenced by solvent quality.
Depending on solvent conditions, chains adopt expanded conformations in good solvents, random-walk statistics under so-called $\theta$-conditions, or compact structures in poor solvents~\cite{de1979scaling,williams1981polymer,khokhlov1994statistical,rubinstein2003polymer}.
Variations in solvent quality affect not only equilibrium chain dimensions but also polymer dynamics through hydrodynamic interactions, which are described in terms of the well known Rouse-Zimm model~\cite{rouse1953theory,zimm1956dynamics,smith1996dynamical}.

Molecular simulations provide valuable microscopic insight into polymer structure and dynamics in solution.
A variety of simulation techniques have been developed to study polymer behavior under flow and at equilibrium.
Brownian dynamics simulations employing bead-spring and bead-rod models have been widely used to investigate polymer stretching under shear flow~\cite{hur2002dynamics,hur2000brownian}, closely reproducing single-molecule DNA experiments~\cite{smith1999single}.
However, such approaches typically rely on implicit solvent models and therefore neglect hydrodynamic interactions.

The importance of hydrodynamic interactions has been emphasized by comparisons between dielectric spectroscopy experiments and Molecular Dynamics (MD) simulations~\cite{kaznessis1999dielectric}.
MD simulations employing implicit solvents fail to capture hydrodynamic coupling and consequently show significant deviations from experimental results.
Explicit-solvent MD approaches, in which hydrodynamic interactions emerge naturally from intermolecular forces, have been proposed to address this issue~\cite{dunweg1991microscopic,pierleoni1991relaxation,dunweg1993molecular,polson2006equilibrium}.
However, the use of Lennard-Jones interactions between solvent particles imposes severe computational constraints, motivating the development of more efficient simulation methods that retain hydrodynamic effects.

Dissipative Particle Dynamics (DPD) is a particle-based simulation technique that provides an efficient description of hydrodynamics in soft-matter systems.
Originally introduced by Hoogerbrugge and Koelman~\cite{hoogerbrugge1992simulating} and later reformulated by Espa\~nol and Warren~\cite{espanol1995statistical}, DPD incorporates soft conservative forces together with dissipative and random pairwise forces.
The latter conserve local momentum and act as a thermostat, ensuring the correct emergence of long-range hydrodynamic interactions~\cite{ripoll2001large}.
DPD has been successfully applied to a wide range of soft-matter and polymer-related problems, which include: biopolymers, surfactants, emulsions, and related systems~\cite{groot1997dissipative,espanol2017perspective,singh2021photo}, as well as phase separation in block copolymers~\cite{groot1999role}, nanocomposites~\cite{laradji2004nanospheres}, lipid bilayers~\cite{laradji2004dynamics}, and flow through polymer brushes~\cite{wijmans2002simulating,huang2006flow}.

In this work, we use DPD simulations to investigate the structural and dynamical properties of dilute solutions of long linear polymers and unknotted, non-concatenated ring polymers, with particular emphasis on the role of hydrodynamic interactions and solvent quality.
The explicit treatment of the solvent allows hydrodynamic coupling to emerge naturally, enabling a direct comparison between systems characterized by different solvent conditions and degrees of inter-polymer interaction.

While pronounced topological effects have been extensively documented for ring polymers in concentrated systems, their relative importance in dilute solutions, particularly in the presence of hydrodynamic interactions and varying solvent quality, remains less clear.
Here, we therefore examine to what extent polymer topology affects chain conformations and dynamics compared with solvent-mediated interactions.
Linear- and ring-polymer systems are systematically compared under both good and poor solvent conditions.

Structural and dynamical observables, including chain mean-square displacements, monomer contacts, and radial distribution functions, are analyzed to characterize conformational organization, transport properties, and the onset of polymer aggregation.
The results clarify the relative roles of solvent quality, hydrodynamic interactions, and chain topology in determining the behavior of polymer solutions.

The remainder of the paper is organized as follows.
Sec.~\ref{Method} describes the model and simulation details.
Results are presented and discussed in Sec.~\ref{NR}.
Conclusions are summarized in Sec.~\ref{Sum}.

%%%
\section{Methodology}\label{Method}
%%%

%%%
\subsection{Dissipative Particle Dynamics (DPD)}\label{ss2a}
%%%
Dissipative Particle Dynamics (DPD) is a coarse-grained simulation method based on interacting beads, where each bead represents a cluster of atoms or molecules.
The coarse-graining inherent to the DPD approach enables simulations over larger length and time scales compared to classical molecular dynamics (MD) simulations~\cite{groot2006local,nikunen2003would,nikunen2007reptational}.
In particular, DPD simulations can reliably capture linear dimensions up to approximately $100$~nm and access timescales extending up to tens of microseconds~\cite{groot1998dynamic,singh2018role,singh2021photo}.

Here, Newton's equations of motion are used to obtain the time-dependent evolution of the system~\cite{groot1997dissipative}:
\begin{equation}\label{eq2}
\frac{d {\vec p}_i}{dt} = \vec{f}_i \, ,
\end{equation}
where ${\vec p}_i = m_i {\vec v}_i$ is the linear momentum, and ${\vec r}_i$ and ${\vec v}_i = d{\vec r}_i/dt$ denote the position and velocity of the $i$-th bead, respectively.
As we treat homopolymers, we take beads of equal mass $m_i = m$ and fix $m=1$ so the value of the total force ${\vec f}_i$ acting on the $i$-th bead is equivalent to its acceleration.
The force term ${\vec f}_i$ consists of three pairwise additive components:
\begin{equation}\label{eq3}
{\vec f}_i = \sum_{i \neq j} \left[ {\vec F}_{ij}^C + {\vec F}_{ij}^D + {\vec F}_{ij}^R \right] \, ,
\end{equation}
where ${\vec F}_{ij}^{C}$ is the conservative force derived from the pairwise interactions between particles, while the terms ${\vec F}_{ij}^D$ and ${\vec F}_{ij}^R$ represent the dissipative and random forces, respectively.
Importantly, the dissipative term depends on the relative velocity between particles while the random force arises due to thermal fluctuations in the system~\cite{espanol2017perspective}.

In DPD simulations, a soft repulsive interaction is employed through the commonly used conservative force between the $i$-th and $j$-th beads~\cite{groot1997dissipative}:
\begin{equation}\label{eq4}
\vec{F}_{ij}^C=
\begin{cases}
a_{ij}\left[1 - \left(r_{ij}/r_c\right) \right]\hat{{r}}_{ij}, & r_{ij} < r_c \\
0, & r_{ij} \geq r_c
\end{cases} \, ,
\end{equation}
where $a_{ij}$ represents the maximum repulsive interaction between the beads.
The values of $a_{ij}$ are chosen carefully for polymer systems, as they strongly influence the structural and dynamical properties of the system~\cite{groot1997dissipative}.
The parameter $r_c$ denotes the cut-off distance for the interaction.
The vector ${\vec r}_{ij} = {\vec r}_i - {\vec r}_j$ represents the relative position between beads $i$ and $j$, $r_{ij} = \lvert {\vec r}_{ij} \rvert$ is the corresponding inter-bead spatial distance and the unit vector ${\hat r}_{ij} = {\vec r}_{ij} / r_{ij}$ denotes the direction of the force acting between the beads.

Finally, the dissipative and random forces have the following forms~\cite{groot1997dissipative}:
\begin{eqnarray}
{\vec F}_{ij}^D & = & -\gamma_D \, \omega^D(r_{ij}) \, ({\hat r}_{ij} \cdot {\vec v}_{ij}) \, {\hat r}_{ij} \, , \label{eq5} \\
{\vec F}_{ij}^R & = & \sigma_R \, \omega^R(r_{ij}) \, \xi_{ij} \, {\hat r}_{ij} \, . \label{eq6}
\end{eqnarray}
Here, $\gamma_D$ and $\sigma_R$ denote the strengths of the dissipative and random forces, corresponding to the friction and noise coefficients, respectively.
Again the unit vector ${\hat r}_{ij}$ defines the direction of these forces, while ${\vec v}_{ij} = {\vec v}_i - {\vec v}_j$ denotes the relative velocity between the $i$-th and $j$-th beads.
The quantity $\xi_{ij}$ is a Gaussian-distributed random number with zero mean and unit variance~\cite{groot1997dissipative, espanol1995statistical, espanol2017perspective}, {\it i.e.}
\begin{eqnarray}\label{eq7}
\langle \xi_{ij}(t) \rangle & = & 0 \, , \\
\langle \xi_{ij}(t) \, \xi_{kl}(t') \rangle & = & ( \delta_{ik} \delta_{jl} + \delta_{il} \delta_{jk} ) \delta(t-t') \, ,
\end{eqnarray}
together with the symmetry relation $\xi_{ij} = \xi_{ji}$ which ensures linear momentum conservation~\cite{groot1997dissipative,espanol1995statistical,espanol2017perspective,singh2021photo}.
As a consequence DPD naturally incorporates flow fields, preserving the correct hydrodynamic behavior of polymer systems consisting of a few hundred beads~\cite{groot1997dissipative,groot1998dynamic,espanol1995statistical}.
Moreover, the system obeys the fluctuation-dissipation relation, ensuring convergence to the correct canonical equilibrium state~\cite{espanol1995statistical,espanol2017perspective}:
\begin{equation}\label{eq9}
\sigma_R^2 = 2\gamma_D k_B T \, , %/ m \, ,
\end{equation}
where $k_B$ and $T$ represent the Boltzmann constant and the equilibrium temperature of the system, respectively.
In Eqs.~\eqref{eq5} and~\eqref{eq6}, $\omega^D(r_{ij})$ and $\omega^R(r_{ij})$ denote the weight functions for the dissipative and stochastic forces, respectively.
These functions are coupled through the following relation~\cite{groot1997dissipative}:
\begin{equation}\label{eq10} 
\omega^D(r_{ij})=\left[\omega^R(r_{ij})\right]^2 = 
\begin{cases} 
\left[1 - \left(r_{ij}/r_c\right)\right]^2, & r_{ij} < r_c \\
0, & r_{ij} \geq r_c
\end{cases} 
\end{equation}
Usually, the weight function $\omega^R(r_{ij})$ is chosen to have a form similar to that of the conservative force $\vec{F}_{ij}^C$ in Eq.~\eqref{eq4}.
However, other functional forms of the weight functions may also be employed, provided they satisfy the constraints imposed by Eqs.~\eqref{eq9} and~\eqref{eq10}~\cite{groot1997dissipative,singh2021photo}.

In the modeling of polymer chains, we typically incorporate two additional potentials.
The first one is the finite-extensible-nonlinear-elastic (FENE) potential~\cite{kremer1990dynamics}:
\begin{equation}\label{bondpot}
E_b = -\frac12 \kappa_b \, R_0^2 \, \ln\!\left[ 1-\left(\frac{r_{ij}}{R_0}\right)^2 \right] \, ,
\end{equation}
which acts between two successive monomers in a polymer chain.
Here, $\kappa_b = 40k_BT / r_c^2$ denotes the bond strength parameter (spring constant) while $R_0 = 2.0 \, r_c$ represents the maximum allowable bond extension.
The second potential is for modeling chain stiffness and it is given by:
\begin{equation}\label{angpot}
 E_a = \frac12 \kappa_a (\cos\theta - \cos\theta_0)^2 \, ,
\end{equation}
where $\kappa_a$ denotes the strength of the potential, while the angle $\theta$ represents the angle between two successive bonds along the polymer chain and $\theta_0$ is the equilibrium value~\cite{nikunen2003would,nikunen2007reptational,singh2021photo,junghans2008transport,hsu2016static,plimpton1995fast}. 
In this study, we adopt the common choices $\kappa_a = 5k_BT$ and $\theta_0 = 180^{\circ}$ for all angles of our polymer chains~\cite{junghans2008transport,hsu2016static,singh2016tailoring,singh2017photo,nikunen2007reptational}. 

In DPD simulations, soft-core interactions allow for a finite overlap between particles.
As a result, unphysical bond crossing between polymer chains may occur.
To suppress such bond-crossing, we employ the {\it modified segmental repulsion potential} (mSRP)~\cite{sirk2012enhanced}.
In this approach, each bond is treated as a {\it fictitious bead} and a soft repulsive interaction is applied between these bonded segments~\cite{sirk2012enhanced}:
\begin{equation}\label{mSRPpot}
{\vec F}_{ij}^S = \kappa_s \left[ 1 - \left(r_{ij} / r_c^s \right)\right] {\hat r}_{ij} \, , \; \text{for} \; r_{ij} < r_c^s \, .
\end{equation}
These fictitious bond beads are assigned at the beginning of the simulation and act as markers for the bond positions.
This approach allows the construction of neighbor lists and the calculation of pairwise interactions in the same manner as for real beads.
The force constant is fixed at $\kappa_s = 100k_BT / r_c$, and the corresponding cut-off distance is chosen as $r_c^s = 0.8r_c$ in reduced DPD units~\cite{singh2023phase,singh2023additive,singh2023amph}.
The inclusion of this short-range repulsive interaction efficiently eliminates unphysical bond-crossing between polymer chains.
All simulations are carried out using the LAMMPS simulation package~\cite{plimpton1995fast}, with the mSRP formulation~\cite{sirk2012enhanced} implemented to integrate the equations of motion.

%%%
\subsection{Model parameters}\label{mpsd}
%%%
We integrate Newton's equations of motion using the {\it velocity-Verlet} algorithm~\cite{groot1997dissipative, plimpton1995fast} to obtain the time evolution of the system.
The simulations are performed using reduced DPD units, where the fundamental quantities are defined as $r_c = m = k_B T = 1.0$~\cite{groot1997dissipative}.
Here, $r_c$ denotes the characteristic interaction length scale (cut-off radius) of the DPD particles, $m$ is the mass, and $k_B T$ represents the characteristic energy scale (see Sec.~\ref{ss2a} and Ref.~\cite{groot1997dissipative}).
The integration time step between two successive iterations is chosen as $\Delta t = 0.01\tau$, where the characteristic time scale is defined as $\tau = (m r_c^2 / k_B T)^{1/2} = 1$~\cite{singh2018role}.
The number density of the system is fixed at $\rho = 3 r_c^{-3}$, which is a commonly adopted value for DPD simulations of polymer liquids~\cite{groot1997dissipative,espanol1995statistical}.
The choice of the friction coefficient $\gamma_D$ is crucial for ensuring rapid equilibration of the system~\cite{singh2017photo, singh2016tailoring}.
In this work, we set $\gamma_D = 4.5(mk_B T / r_c^2)^{1/2}$, a widely used value for polymer systems~\cite{singh2017photo,singh2016tailoring}.
At this value, the dissipative force effectively captures hydrodynamic interactions while maintaining numerical stability over the simulation time steps~\cite{groot1997dissipative,hashimoto1986late}.
Nevertheless, different choices of $\gamma_D$ may be appropriate depending on the system under investigation and the desired level of coarse-graining~\cite{espanol1995statistical,groot1997dissipative}.
All simulations are carried out at fixed temperature $T = 1$ in reduced DPD units, assuming for reference standard room temperature ($=297$ kelvin~\cite{singh2023amph,singh2023additive,sam2024emulsion,sam2022brush}).

The repulsive interaction strength $a_{ij}$ plays an important role in determining both the static and dynamical properties of multicomponent systems~\cite{groot1998dynamic,singh2018role}.
Proper tuning of this parameter is therefore essential for accurately modeling complex fluids within the DPD framework~\cite{groot1997dissipative}.
In the present work we investigate solutions of linear and ring polymers, and consider solvents of different qualities.
For bead pairs of the same type, the conservative interaction parameter is typically chosen as $a_{ii} = 25k_B T / r_c$~\cite{groot1997dissipative,singh2016tailoring,singh2017photo}.
For bead pairs of different type, the interaction strength is expressed as $a_{ij} = a_{ii} + 3.27\chi_{ij}$~\cite{groot1997dissipative,singh2023amph,singh2023additive} where the $\chi_{ij}$'s are properly tuned Flory-Huggins-like~\cite{rubinstein2003polymer} interaction parameters.
In this study, we set $a_{ss} = 25k_B T / r_c$ and $a_{pp} = 30k_B T / r_c$ for solvent-solvent and polymer-polymer bead pairs, respectively.
Then:
\begin{itemize}
\item
For good solvent conditions, the polymer-solvent interaction is made more favorable than the polymer-polymer interaction by choosing $a_{ps} = 25k_B T / r_c$.
\item
For poor solvent conditions, we increased the polymer-solvent interaction parameter to $a_{ps} = 40k_B T / r_c$, thereby promoting effective polymer-polymer attraction.
\end{itemize}
%

%%%
\subsection{Mapping to real units}\label{sec:MappingRealUnits}
%%%
In this study, ten water molecules are coarse-grained into a single DPD bead.
Since the physical properties of water are well known, this mapping allows a direct estimation of the characteristic length, mass, and energy scales of the DPD simulation.
A single water molecule has an effective volume $V_{\rm water} = 30~\text{\AA}^3$ and a mass density $\rho_m = 1~{\rm g} \cdot {\rm cm}^{-3}$ \cite{singh2023additive,singh2023amph}.
Accordingly, the volume associated with one DPD bead is $V_{\rm DPD} = 10V_{\rm water}$.
As anticipated in Sec.~\ref{mpsd}, the number density of the system is set to $\rho = 3r_c^{-3}$, implying that three DPD beads occupy a cubic volume of side length $r_c$.
Equating this volume to the total volume of three DPD beads gives $r_c^3 = 3V_{\rm DPD} = 30V_{\rm water} = 900~\text{\AA}^3$, from which the DPD cut-off length is obtained as $r_c \simeq 0.97~{\rm nm}$.
The corresponding characteristic bead mass is $m = 180~\text{Da}$, consistent with the coarse-grained representation of ten water molecules per DPD bead~\cite{groot1997dissipative,singh2023additive}.
Finally, using these characteristic length, mass, and energy (room temperature $k_B T = 297$ kelvin, see Sec.~\ref{mpsd}) scales, the corresponding dimensional unit of time is estimated as $\tau = (m r_c^2 / k_B T)^{1/2} \simeq 8.2~{\rm ps}$~\cite{groot1997dissipative,singh2023additive,singh2023amph}.
This time scale reflects the accelerated dynamics inherent to DPD simulations arising from the use of soft-core interactions~\cite{symeonidis2006schmidt,junghans2008transport}.

Another realistic estimate of the physical time scale can be obtained by calibrating the diffusion coefficient from DPD simulations against experimental measurements~\cite{Kremer1990,Wim2018,Yong2013}.
In the simulations, the diffusion coefficient is typically of the order $D_{\rm sim} \sim 10^{-2}~{\rm nm}^2 / \tau \sim 10^{-9}~{\rm m}^2 / {\rm s}$~\cite{Yong2013}, whereas experimental values are  $D_{\rm exp} \sim 10^{-11}~{\rm m}^2 / {\rm s}$~\cite{nikoobakht2003preparation,Janez2020,hsu2016static}, {\it i.e.} two orders of magnitude lower.
Based on this comparison, a more appropriate DPD time unit for the present system is estimated to be $\tau \sim 1.0~{\rm ns}$~\cite{Wim2018,junghans2008transport,singh2023additive}.
In conclusion, the physical time scale $\tau$ is in the range $10^{-11}$-$10^{-9}$ seconds.

%%%
\subsection{Systems' preparation and simulation details}\label{sec:SimulDetails}
%%%
We consider relatively dilute solutions of linear and circular (ring) polymers in the presence of explicit solvent.
This study compares the static and dynamic properties of these systems, while maintaining a fixed polymer bead concentration of $2\%$ in both cases.
For both systems, we employ a cubic simulation box of side length $L_{\rm box} = 54\,r_c$.
Using the standard DPD number density $\rho = 3\,r_c^{-3}$ (see Sec.~\ref{mpsd}), the total number of particles in the system is $L_{\rm box}^3 \rho = 472,\!392$.
For linear solutions, we take $M=40$ chains composed of $N=250$ monomers each, which fixes the total polymer concentration at approximately $2\%$.
For ring solutions, we consider instead $M=20$ chains and each chain consists of $N=500$ monomers.

Initial conformations are constructed as the following.
First, polymer chains are placed randomly inside the simulation box.
For ring solutions, special care is taken to ensure that the rings neither form knots nor become concatenated with one another.
Then, solvent particles are added at random positions until the nominal number density $\rho = 3r_c^{-3}$ is reached.
The systems are subsequently equilibrated for a sufficiently long time, $t = 5 \times 10^4 \, \tau$, to achieve a uniformly mixed state.
During this equilibration stage, all interaction parameters are chosen to be identical, {\it i.e.}\ $a_{pp} = a_{ss} = a_{ps} = 25\,k_B T/r_c$ (see Sec.~\ref{mpsd}), corresponding to fully compatible polymer--solvent interactions.
This choice ensures a homogeneous, well-mixed system in which all beads experience similar interactions, enabling rapid and efficient equilibration.
After equilibration, the simulation time is reset to $t = 0$, and the system dynamics are monitored for a total runtime of $T = 10^5\,\tau$.
With $\Delta t = 0.01\tau$ (see Sec.~\ref{mpsd}), a total of $10^7$ DPD integration steps were performed.
The simulations were executed using 2 MPI processes and required approximately 1200 hours of wall-clock time, equivalent to 2400 CPU-hours, corresponding to an average computational cost of about $40 \, {\rm seconds} / \tau$.

For both the linear and ring solution and for good and poor solvent conditions, we have generated $3$ independent trajectories where each trajectory starts from a different initial conformation.
In this way we are able to monitor constantly the robustness of our results, demonstrating that they are essentially independent of the assumed initial conformation (see ``Results'' Sec.~\ref{NR}).

%%%
\section{Results}\label{NR}
%%%

%%%
\subsection{Polymer dynamics}\label{sec:Results-PolymerDynamics}
\begin{figure*}
%\centering
\includegraphics[width=1.0\textwidth]{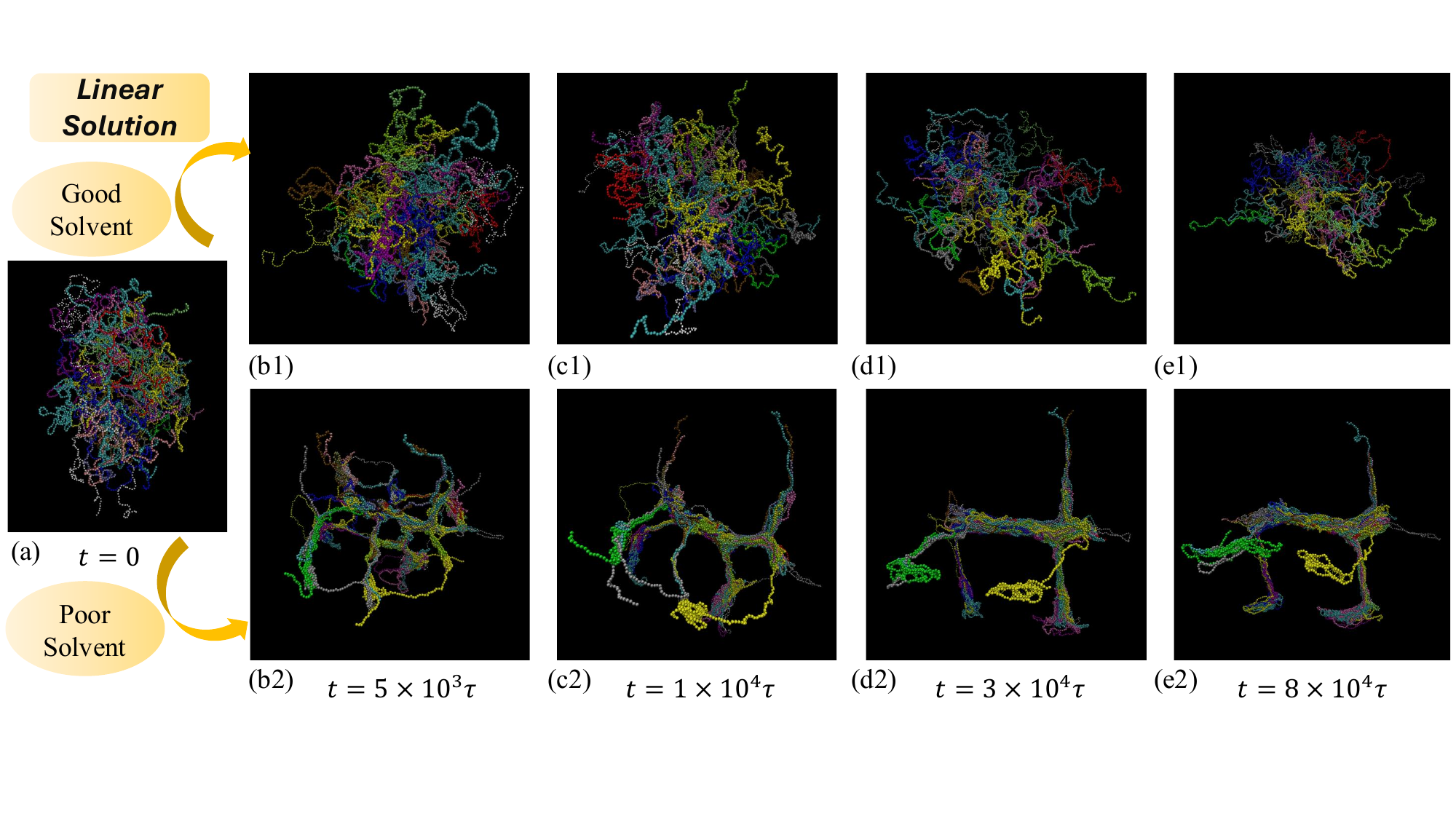}
\caption{
Snapshots illustrating the typical temporal evolution of linear polymer solutions, under good and poor solvent conditions.
(a) Homogeneous conformation at $t = 0$.
(b1)-(e1) Time evolution in good solvent.
(b2)-(e2) Time evolution in poor solvent.
Configurations are sampled at the chosen time steps (in DPD $\tau$-units, see Sec.~\ref{mpsd}) indicated in the figure.
}
\label{Fig1}
\end{figure*}

Fig.~\ref{Fig1} shows representative conformations of solutions of $M=40$ linear polymers of single-chain contour length $N=250$, evolving under good and poor solvent conditions. 
After equilibration in neutral solvent (Fig.~\ref{Fig1}(a); see Sec.~\ref{sec:SimulDetails} for technical details), the same initial configuration is evolved under good (panels (b1)-(e1)) and poor (panels (b2)-(e2)) solvent conditions.
Under good solvent conditions the polymer chains maintain swollen conformations, while in the opposite case of poor solvent local density fluctuations emerge and progressively coarsen into compact aggregates, driving the whole system toward a compact state.
Within the simulated time window, complete collapse into a single compact globule is not observed.
In fact, the collapse dynamics proceed slowly, likely due to the large chain length and hydrodynamic interactions; at longer times, the formation of anisotropic clusters is observed.

\begin{figure*}
%\centering
\includegraphics[width=1.0\textwidth]{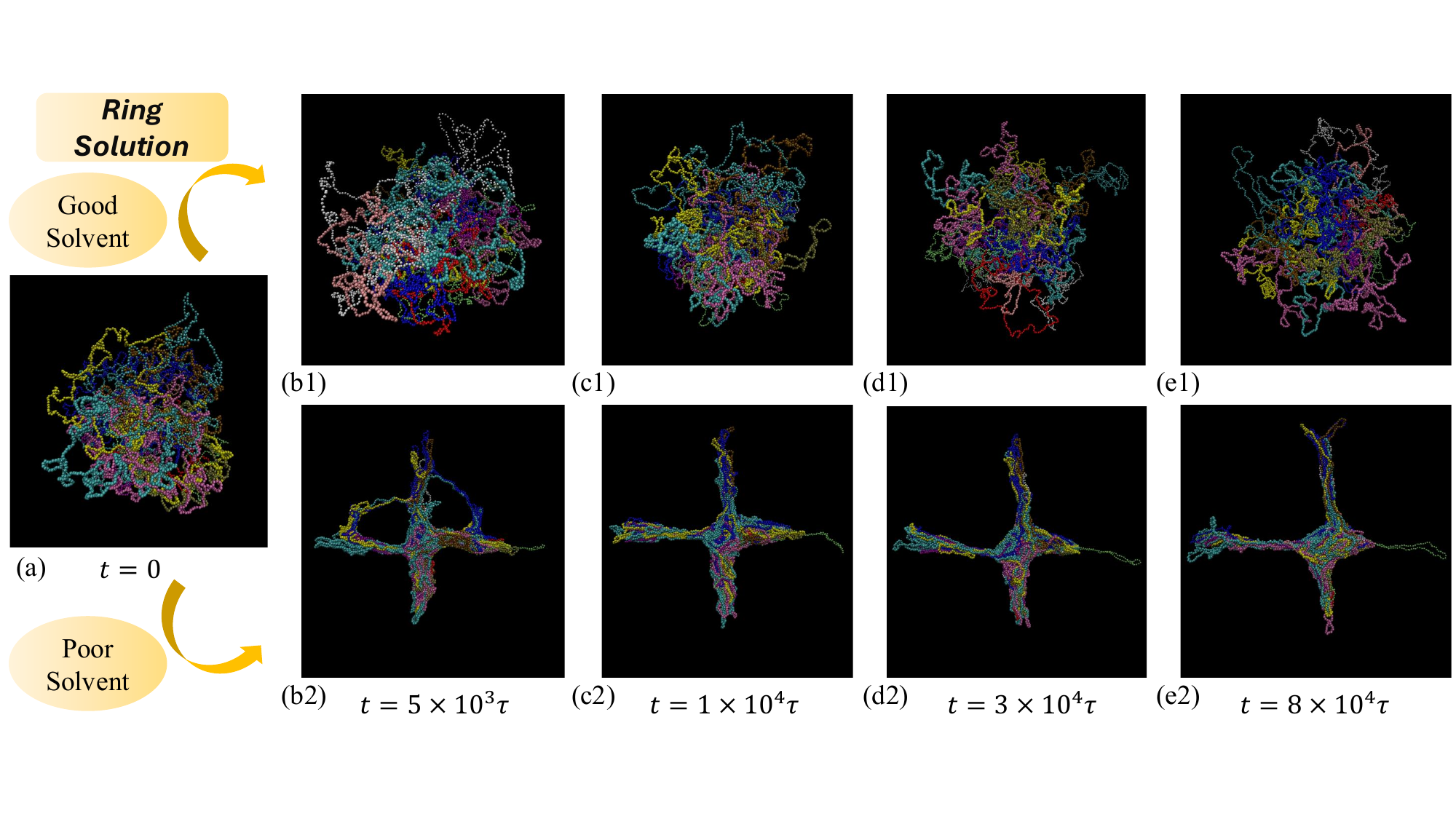}
\caption{
Snapshots illustrating the typical temporal evolution of ring polymer solutions, under good and poor solvent conditions.
The organization of the figure is as in Fig.~\ref{Fig1}.
}
\label{Fig2}
\end{figure*}

Similarly, Fig.~\ref{Fig2} shows representative conformations of solutions of $M=20$ ring polymers of single-chain contour length $N=500$.
From the equilibrated initial configuration shown in Fig.~\ref{Fig2}(a), we consider the time evolution of the system under good (panels (b1)-(e1)) and poor (panels (b2)-(e2)) solvent conditions.
In good solvent conditions, the rings remain swollen and move independently and essentially unconstrained from each other.
Conversely, in poor solvent each ring undergoes collapse, and spatial proximity between chains promotes the formation of dense aggregates.
Within the simulated time window, the system does not evolve into a single compact globule but rather forms partially collapsed, anisotropic clusters.
Importantly, notice that bond crossing is strictly prohibited by the mSRP potential (see technical details in Sec.~\ref{ss2a}, Eq.~\eqref{mSRPpot}), ensuring that no artificial cross-linking or concatenation occurs.

It is interesting to notice that, at the considered dilute conditions, linear and ring systems behave similarly, in marked contrast with the phenomenology observed in dense environments~\cite{RubinsteinPRL1986,kapnistos2008unexpected,RosaPRL2014,MichielettoTurner2016,Ubertini_PRE2021,Ubertini_Macromolecules2023,TubianaPhysRep2024,Schroeder2025}, {\it i.e.} in the near absence of solvent.
To characterize this similarity at the level of chain dynamics more quantitatively, we measure
(i) the mean-square displacement of the chain center of mass,
\begin{equation}\label{eq:DefineG3}
\Delta r_{\rm cm}^2(t) \equiv \frac1{T-t} \int_0^{T-t} dt' \left( \vec r_{\rm cm}(t'+t) - \vec r_{\rm cm}(t') \right)^2 \, ,
\end{equation}
where $\vec r_{\rm cm}(t) = \frac1N \sum_{n=1}^N \vec r_n(t)$ denotes the center-of-mass position of the generic polymer chain at time $t$ and $T$ is the total runtime of a single trajectory (see Sec.~\ref{sec:SimulDetails} for details)
and
(ii) the associated scaling exponent,
\begin{equation}\label{eq:Alpha-t}
\alpha(t) \equiv \frac12 \frac{ \log\left( \Delta r_{\rm cm}^2(t+1) / \Delta r_{\rm cm}^2(t) \right) }{ \log\left( (t+1) / t \right) } \, ,
\end{equation}
quantifying the local slope of $\Delta r_{\rm cm}^2(t)$ as a function of $t$ in log-log representation.
The exponent $\alpha(t)$ characterizes the nature of polymer dynamics: $\alpha = 0.5$ corresponds to normal diffusive motion, whereas $\alpha < 0.5$ indicates sub-diffusive dynamics arising from constraints such as chain connectivity, topological interactions, and molecular crowding in polymer systems~\cite{rubinstein2003polymer}.

\begin{figure*}
%\centering
\includegraphics[width=1.0\textwidth]{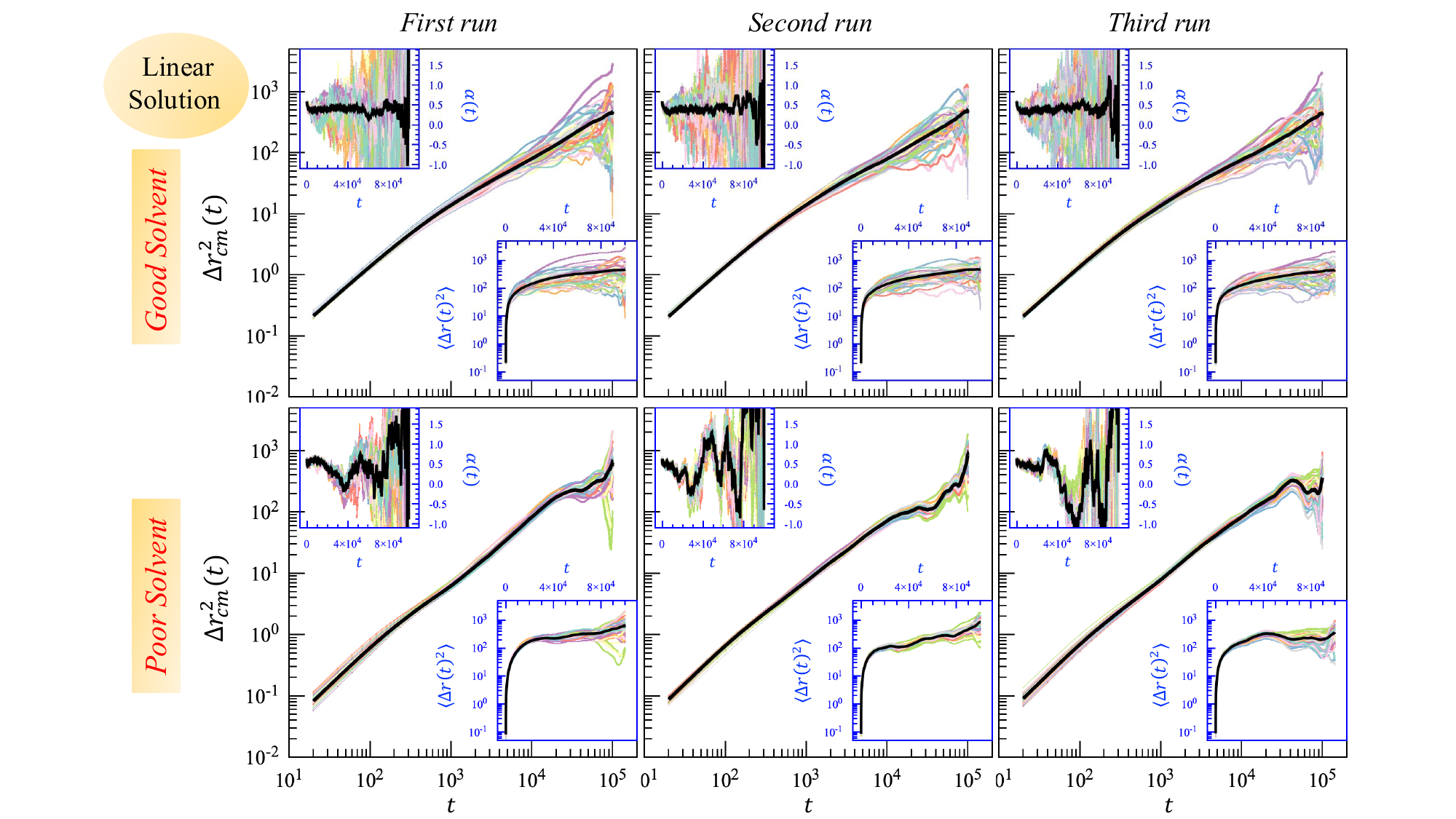}
\caption{
Mean-square displacement, $\Delta r_{\rm cm}^2(t)$ (Eq.~\eqref{eq:DefineG3}), of the center of mass of linear polymers as a function of time $t$.
The first and second rows correspond to good and poor solvent conditions, respectively.
Colored curves represent the distinct polymer chains, while the solid black curve denotes the average over the chains ensemble.
Main plot: log-log representation; lower inset: linear-log representation; upper inset: effective diffusion exponent $\alpha(t)$ (Eq.~\eqref{eq:Alpha-t}).
Different columns are for the 3 simulated independent runs.
}
\label{Fig3}
\end{figure*}
\begin{figure*}
%\centering
\includegraphics[width=1.0\textwidth]{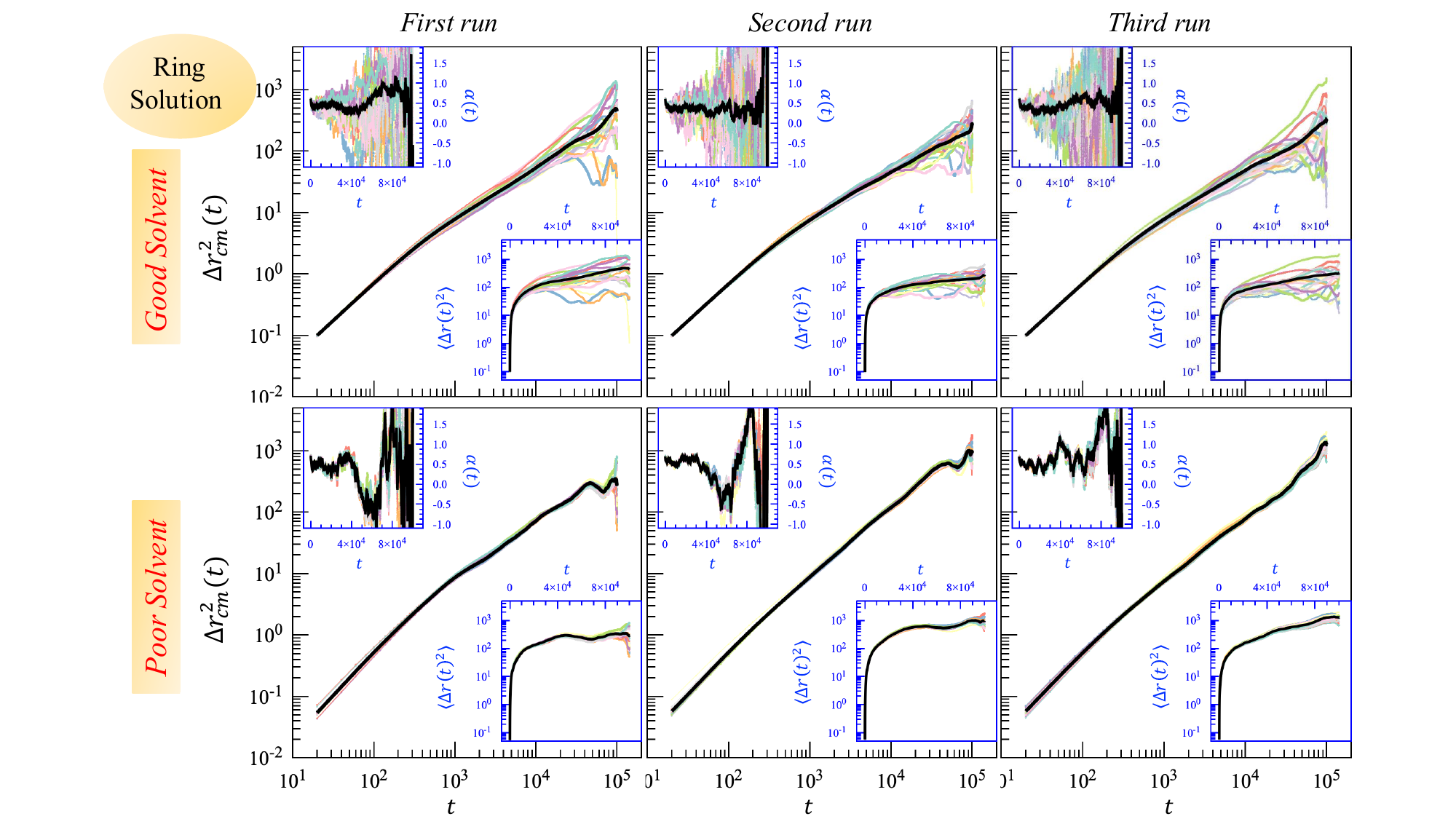}
\caption{
Mean-square displacement, $\Delta r_{\rm cm}^2(t)$ (Eq.~\eqref{eq:DefineG3}), of the center of mass of ring polymers as a function of time $t$.
Notation, colors and symbols are as in Fig.~\ref{Fig3}.
}
\label{Fig4}
\end{figure*}

Figs.~\ref{Fig3} and~\ref{Fig4} show results for $\Delta r_{\rm cm}^2(t)$ and the related $\alpha(t)$ for the individual (linear and ring, respectively, see colored lines) polymers under good and poor solvent conditions, for three independent runs. 
Under good solvent conditions (upper rows in Figs.~\ref{Fig3} and~\ref{Fig4}) the polymers remain well dispersed, diffusing almost independently throughout the solvent, as indicated by the curves' spreading at times $\gtrsim 10^4\tau$ (see also the linear-log representation of the plots in the lower insets).
$\Delta r_{\rm cm}^2(t)$ exhibits a sustained increase with time, indicating persistent chain mobility and the absence of large-scale aggregation. 
Correspondingly (upper insets), the effective diffusion exponent, $\alpha(t)$, fluctuates around the value $0.5$ (black line) expected for normal diffusion. 
In contrast, under poor solvent conditions attractive polymer-polymer interactions strongly suppress chain mobility and promote the formation of dense local clusters (see corresponding panels in Figs.~\ref{Fig1} and~\ref{Fig2}).
As a consequence (lower rows in Figs.~\ref{Fig3} and~\ref{Fig4}, main panels and lower insets), after early-time transient dynamics that is close to that reported for good solvent conditions, chains' diffusion at later times is less heterogenous and the distinct curves for $\Delta r_{\rm cm}^2(t)$ show visibly correlated motion, and a plateau-like regime indicating strongly hindered center-of-mass motion associated with partially collapsed anisotropic structures.
Consequently, the effective diffusion exponents $\alpha(t)$ (upper insets) take values significantly below the diffusive limit, signaling the emergence of subdiffusive dynamics, accompanied by stronger fluctuations due to slow structural rearrangements and constrained motion within the partially collapsed polymer aggregates.

Again, it is interesting to notice that linear and ring polymers behave alike.

%%%
\subsection{Polymer conformations}\label{sec:Results-PolymerConformations}
We now turn our attention to the impact of solvent quality on the spatial organization of the solution.

\begin{figure}
%\centering
\includegraphics[width=0.60\textwidth]{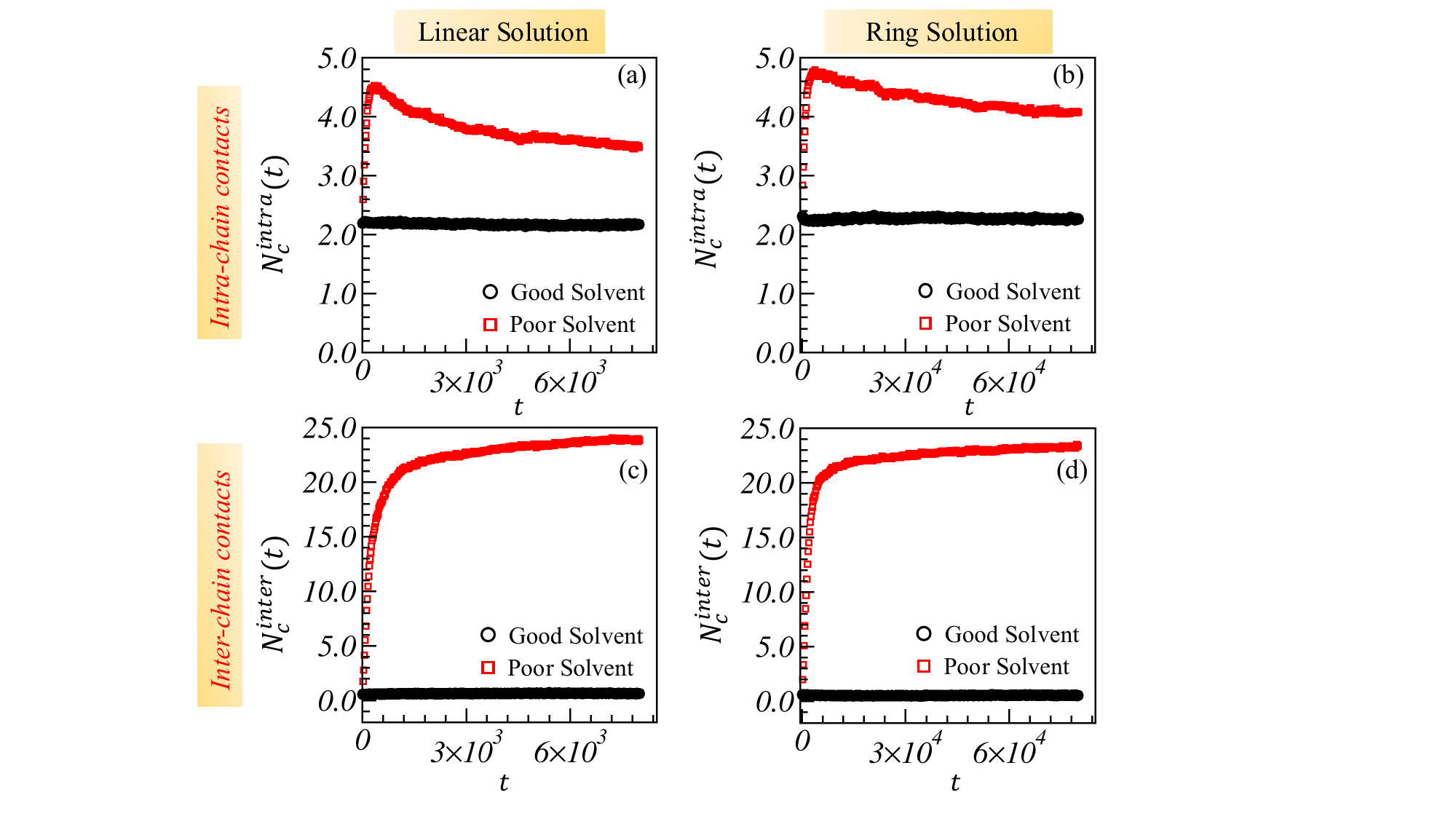}
\caption{
Time evolution of intra-chain contacts per monomer $N_c^{\rm intra}(t)$ (Eq.~\eqref{eq:Contacts-IntraChain}, panels (a) and (b)) and inter-chain contacts per monomer $N_c^{\rm inter}(t)$ (Eq.~\eqref{eq:Contacts-InterChain}, panels (c) and (d)), for linear and ring solutions in good and poor solvent conditions.
}
\label{Fig5}
\end{figure}

First, we consider how chains' compaction evolve over time.
Denoting with $\vec r_n^{\,m}$ the spatial position of the $n$-th monomer of the $m$-th chain, we measure the mean number of intra-chain contacts per monomer,
\begin{align}\label{eq:Contacts-IntraChain}
& N_c^{\rm intra}(t) \equiv \nonumber\\
& \frac1{M N(N-1)} \sum_{m=1}^M \sum_{n \neq n'=1}^{N} \Theta( r_{\rm co} - |\vec r_n^{\,m}(t) - \vec r_{n'}^{\,m}(t)| ) \, ,
\end{align}
and the mean number of inter-chain contacts per monomer,
\begin{align}\label{eq:Contacts-InterChain}
& N_c^{\rm inter}(t) \equiv \nonumber\\
& \frac1{M(M-1) N^2} \sum_{m\neq m'=1}^M \sum_{n,n'=1}^N \Theta( r_{\rm co} - |\vec r_n^{\,m}(t) - \vec r_{n'}^{\,m'}(t)| ) \, ,
\end{align}
as a function of time $t$.
In Eqs.~\eqref{eq:Contacts-IntraChain} and~\eqref{eq:Contacts-InterChain}, $\Theta(x)$ denotes the conventional Heaviside step function counting the monomer pairs located within the conventional cut-off distance $r_{\rm co} = 2r_c$ ({\it i.e.} two monomer units) from each other.
Results are shown in Fig.~\ref{Fig5}.
For both linear and ring polymer systems, and under good solvent conditions, local packing remains low and essentially time-independent, with $\langle N_c^{\rm intra} \rangle \approx 2-3$ and negligible inter-chain contacts.
Instead, under poor solvent condition, intra-chain contacts initially dominate but are progressively complemented by a growing inter-chain contribution, reflecting enhanced inter-chain association and collective compaction.

\begin{figure}
%\centering
\includegraphics[width=0.60\textwidth]{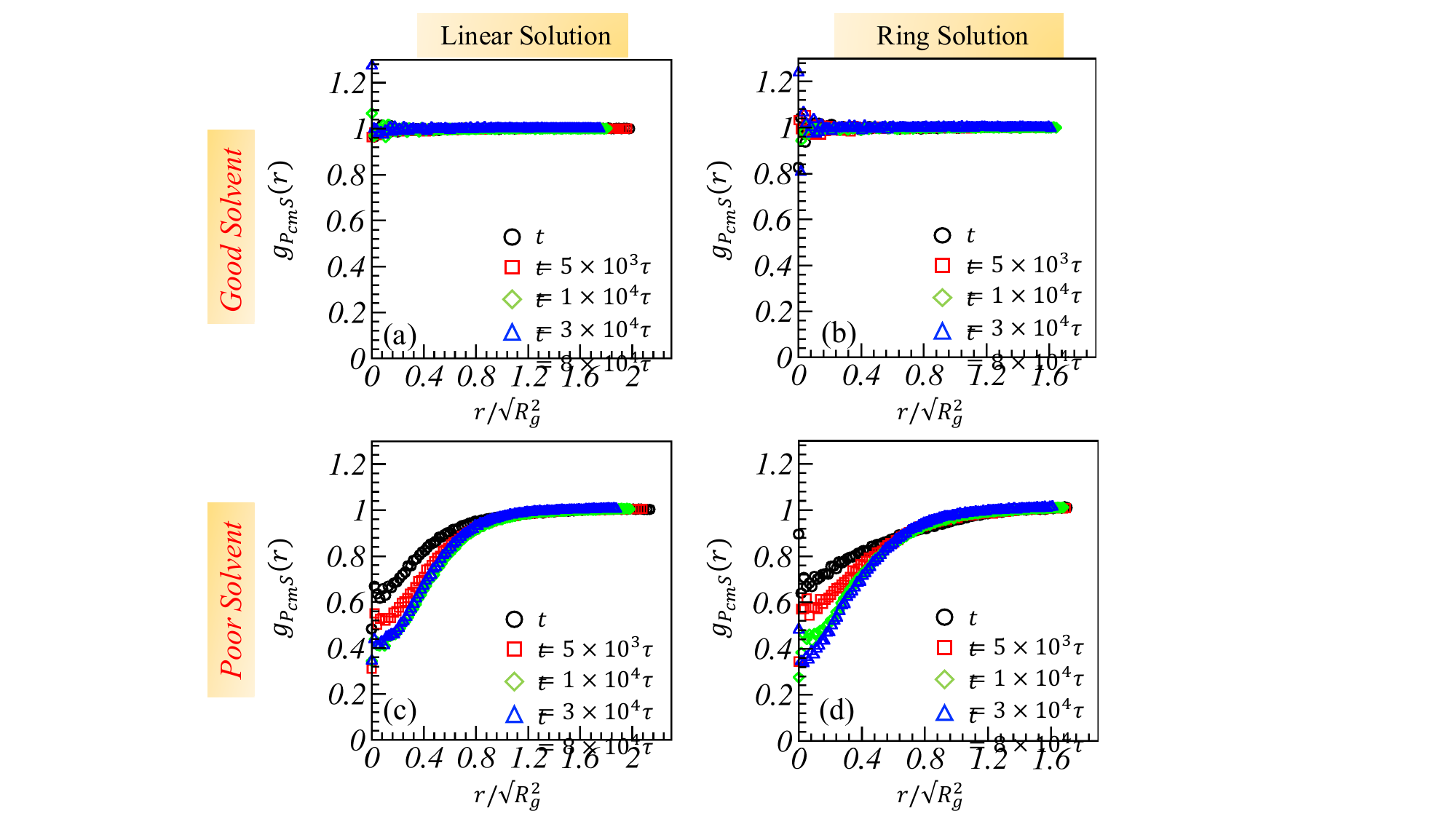}
\caption{
Radial distribution function of solvent particles relative to the polymer center of mass, $g_{p_{\rm cm} \, s}(r)$, taken at the same representative time steps as in Figs.~\ref{Fig1} and~\ref{Fig2}.
Distances are normalized by the root-mean-square gyration radius of the chain at time $t$, $\sqrt{R_g^2(t)}$ (see Eq.~\eqref{eq:<Rg2>}).
}
\label{Fig6}
\end{figure}

To explore in more detail the time behavior of the polymers in relation to that of the surrounding solvent, we measured the radial distribution function $g_{p_{\rm cm} \, s}(r)$ of solvent particles as a function of spatial distance $r$ from the polymer center of mass and at the same representative time steps as in Figs.~\ref{Fig1} and~\ref{Fig2}. 
Results are shown in Fig.~\ref{Fig6}, where -- to facilitate the comparison between different instants of the time evolution of the polymers (this is especially important under poor solvent conditions) -- the $x$-axis is expressed in units of the root-mean-square gyration radius of the chains at time $t$,
\begin{equation}\label{eq:<Rg2>}
\sqrt{ R_g^2(t) } \equiv \sqrt{ \frac1{M N} \sum_{m=1}^M \sum_{n=1}^N \left( \vec r_n^{\,m}(t) - \vec r_{\rm cm}^{\,m}(t) \right)^2 } \, ,
\end{equation}
where $\vec r_{\rm cm}^{\,m}(t) = \frac1N \sum_{n=1}^N \vec r_n^{\,m}(t)$ is the time-dependent spatial position of the centre of mass of the $m$-th chain.
In good solvent conditions, solvent particles remain accessible near the chain center of mass at all times, consistent with a stable swollen state.
In poor solvent instead, $g_{p_{\rm cm} \, s}(r)$ decreases near the origin as solvent is progressively expelled from the collapsing polymer.

\begin{figure}
%\centering
\includegraphics[width=0.60\textwidth]{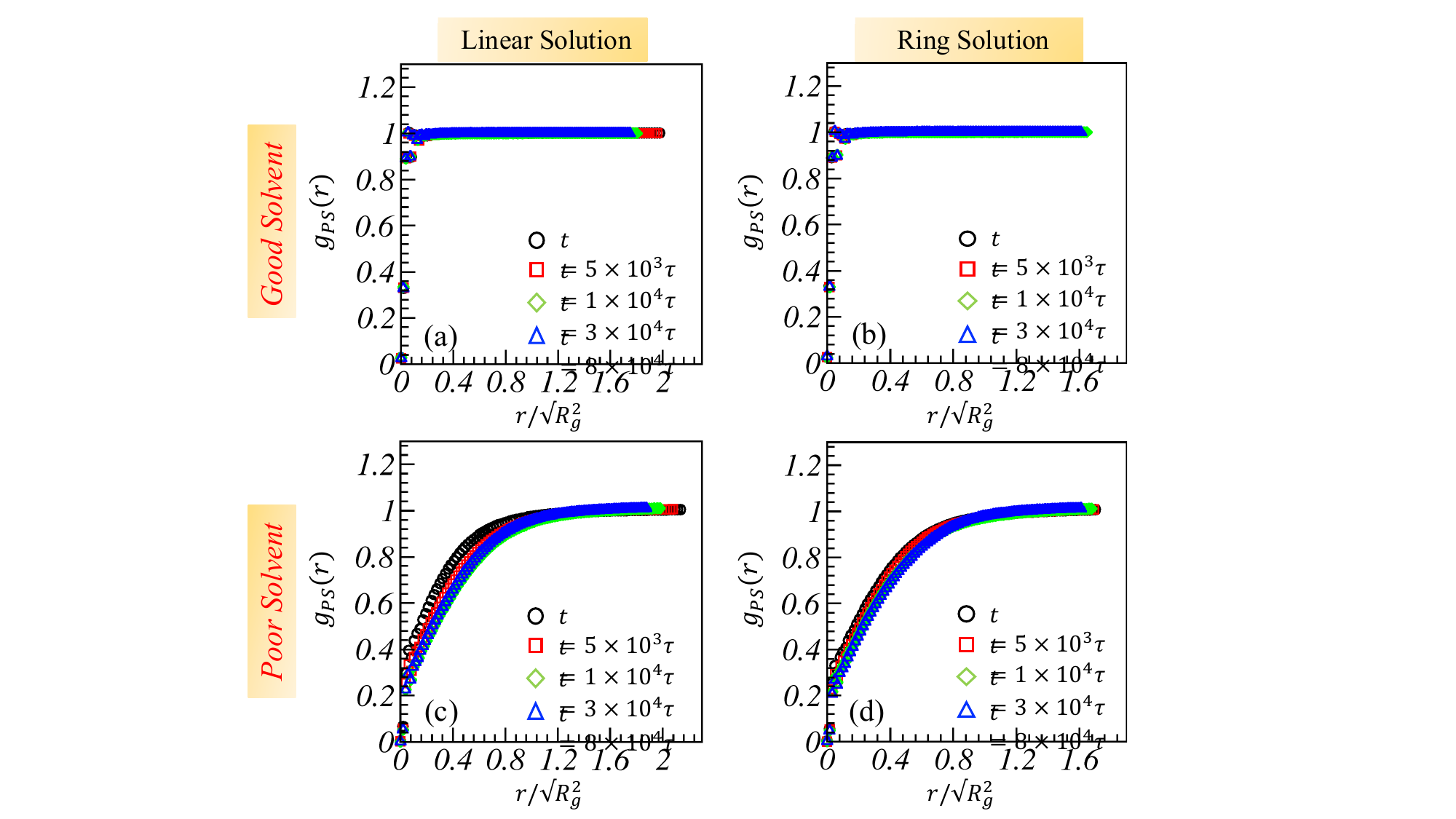}
\caption{
Radial distribution function of solvent particles around a reference monomer particle, $g_{ps}(r)$.
Notation, colors and symbols are as in Fig.~\ref{Fig6}.
}
\label{Fig7}
\end{figure}

In agreement with these findings, the radial distribution function $g_{ps}(r)$ of solvent particles as a function of the spatial distance from a reference monomer particle confirms (see Fig.~\ref{Fig7}) that in good solvent the local solvent environment remains essentially unchanged.
In poor solvent instead, $g_{ps}(r)$ is suppressed at short distances, reflecting reduced solvent accessibility in the collapsed state.

\begin{figure}
%\centering
\includegraphics[width=0.60\textwidth]{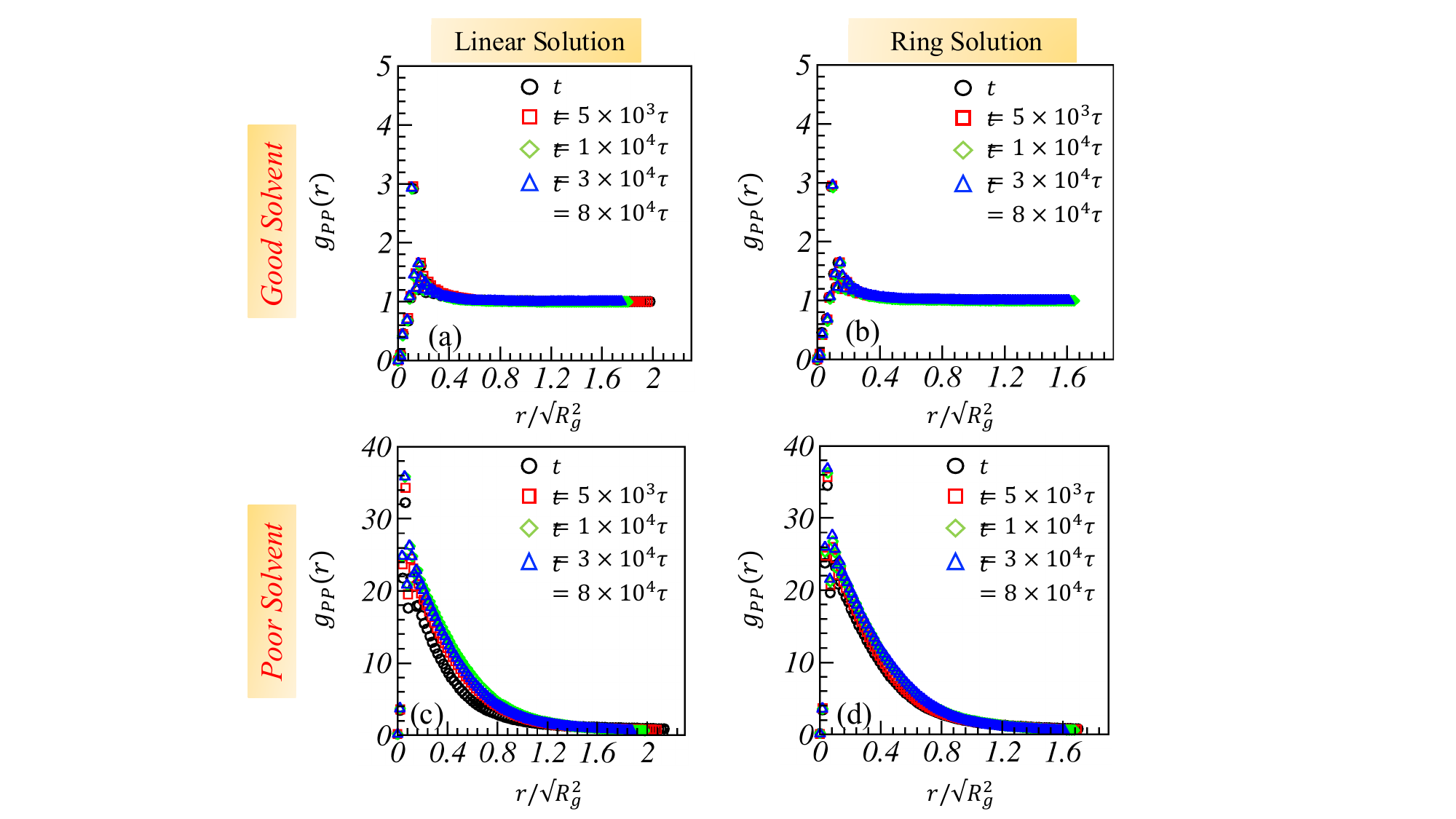}
\caption{
Monomer-monomer radial distribution function, $g_{pp}(r)$.
Notation, colors and symbols are as in Fig.~\ref{Fig6}.
}
\label{Fig8}
\end{figure}

Finally, the monomer-monomer radial distribution function $g_{pp}(r)$ exhibits (see Fig.~\ref{Fig8}) the expected connectivity-induced peaks at short distances and approaches unity at larger separations in good solvent.
In poor solvent, enhanced short-range correlations and elevated peak amplitudes reflect the formation of dense, compact structures, with $g_{pp}(r)$ decaying toward unity beyond the effective globule size.

Notice that these results hold for linear and ring solutions alike, and that the agreement remains quantitative, {\it i.e.} chains' topology plays no significant role on the length and time scales considered in this work.

%%%
\section{Summary and Conclusion}\label{Sum}
%%%
In this work, Dissipative Particle Dynamics simulations were used to investigate the influence of hydrodynamic interactions, solvent quality, and polymer topology on the structural and dynamical behavior of dilute solutions of linear and unknotted, non-concatenated ring polymers.

The simulations reveal that solvent quality is the dominant factor controlling both polymer conformation and dynamics.
Under good solvent conditions, linear and ring polymers remain swollen and well dispersed, preserving substantial solvent accessibility within the polymer volume, as confirmed by radial distribution functions and contact analyses (Figs.~\ref{Fig5}, \ref{Fig6}, \ref{Fig7} and~\ref{Fig8}).
In this regime, both architectures display nearly diffusive center-of-mass motion throughout the simulated time window (Figs.~\ref{Fig3} and~\ref{Fig4}), indicating sustained chain mobility and weak inter-chain interactions.

Upon switching to poor solvent conditions, polymer-polymer attraction drives the formation of dense and irregular aggregates.
Rather than undergoing independent chain collapse, polymers become incorporated into partially collapsed clusters characterized by significant interpenetration and crowding. 
These effects hinder large-scale structural rearrangements, leading to kinetically trapped states and a pronounced slowdown of translational dynamics.
Consistently, the mean-square displacement develops extended plateau regions and the effective diffusion exponent decreases below the value expected for normal diffusion, displaying in particular notable fluctuations over time.

A key result of this study is that, despite the distinct molecular architectures, linear and ring polymers exhibit qualitatively similar behavior under both solvent conditions.
The transition from dispersed to aggregated states, as well as the associated dynamical slowdown, is primarily governed by solvent-mediated interactions, while chain topology plays only a minor role within the range of concentrations and chain lengths considered here.
These findings suggest that, in dilute solutions where hydrodynamic interactions are present, solvent quality largely outweighs topological effects in determining the collective structural and dynamical properties of the system.

%%%
\section*{Data availability}
%%%
Data will be made available upon reasonable request.

%%%
\section*{Conflicts of interest}
%%%
There are no conflicts of interest to declare.

%%%
\section*{Acknowledgments}
%%%
AKS acknowledges SISSA high-performance computing cluster infrastructure (Ulysses), where simulations for this paper were made.
%AR acknowledges financial support from the PNRR Grant CN00000013 (CN-HPC, M4C2I1.4, Spoke 7), funded by Next Generation EU.

%\clearpage
%\newpage

\bibliography{mybib}

\begin{thebibliography}{62}%
\makeatletter
\providecommand \@ifxundefined [1]{%
 \@ifx{#1\undefined}
}%
\providecommand \@ifnum [1]{%
 \ifnum #1\expandafter \@firstoftwo
 \else \expandafter \@secondoftwo
 \fi
}%
\providecommand \@ifx [1]{%
 \ifx #1\expandafter \@firstoftwo
 \else \expandafter \@secondoftwo
 \fi
}%
\providecommand \natexlab [1]{#1}%
\providecommand \enquote  [1]{``#1''}%
\providecommand \bibnamefont  [1]{#1}%
\providecommand \bibfnamefont [1]{#1}%
\providecommand \citenamefont [1]{#1}%
\providecommand \href@noop [0]{\@secondoftwo}%
\providecommand \href [0]{\begingroup \@sanitize@url \@href}%
\providecommand \@href[1]{\@@startlink{#1}\@@href}%
\providecommand \@@href[1]{\endgroup#1\@@endlink}%
\providecommand \@sanitize@url [0]{\catcode `\\12\catcode `\$12\catcode
  `\&12\catcode `\#12\catcode `\^12\catcode `\_12\catcode `\%12\relax}%
\providecommand \@@startlink[1]{}%
\providecommand \@@endlink[0]{}%
\providecommand \url  [0]{\begingroup\@sanitize@url \@url }%
\providecommand \@url [1]{\endgroup\@href {#1}{\urlprefix }}%
\providecommand \urlprefix  [0]{URL }%
\providecommand \Eprint [0]{\href }%
\providecommand \doibase [0]{https://doi.org/}%
\providecommand \selectlanguage [0]{\@gobble}%
\providecommand \bibinfo  [0]{\@secondoftwo}%
\providecommand \bibfield  [0]{\@secondoftwo}%
\providecommand \translation [1]{[#1]}%
\providecommand \BibitemOpen [0]{}%
\providecommand \bibitemStop [0]{}%
\providecommand \bibitemNoStop [0]{.\EOS\space}%
\providecommand \EOS [0]{\spacefactor3000\relax}%
\providecommand \BibitemShut  [1]{\csname bibitem#1\endcsname}%
\let\auto@bib@innerbib\@empty
%</preamble>
\bibitem [{\citenamefont {Rubinstein}(1986)}]{RubinsteinPRL1986}%
  \BibitemOpen
  \bibfield  {author} {\bibinfo {author} {\bibfnamefont {M.}~\bibnamefont
  {Rubinstein}},\ }\href {https://doi.org/10.1103/PhysRevLett.57.3023}
  {\bibfield  {journal} {\bibinfo  {journal} {Phys. Rev. Lett.}\ }\textbf
  {\bibinfo {volume} {57}},\ \bibinfo {pages} {3023} (\bibinfo {year}
  {1986})}\BibitemShut {NoStop}%
\bibitem [{\citenamefont {Kapnistos}\ \emph {et~al.}(2008)\citenamefont
  {Kapnistos}, \citenamefont {Lang}, \citenamefont {Vlassopoulos},
  \citenamefont {Pyckhout-Hintzen}, \citenamefont {Richter}, \citenamefont
  {Cho}, \citenamefont {Chang},\ and\ \citenamefont
  {Rubinstein}}]{kapnistos2008unexpected}%
  \BibitemOpen
  \bibfield  {author} {\bibinfo {author} {\bibfnamefont {M.}~\bibnamefont
  {Kapnistos}}, \bibinfo {author} {\bibfnamefont {M.}~\bibnamefont {Lang}},
  \bibinfo {author} {\bibfnamefont {D.}~\bibnamefont {Vlassopoulos}}, \bibinfo
  {author} {\bibfnamefont {W.}~\bibnamefont {Pyckhout-Hintzen}}, \bibinfo
  {author} {\bibfnamefont {D.}~\bibnamefont {Richter}}, \bibinfo {author}
  {\bibfnamefont {D.}~\bibnamefont {Cho}}, \bibinfo {author} {\bibfnamefont
  {T.}~\bibnamefont {Chang}},\ and\ \bibinfo {author} {\bibfnamefont
  {M.}~\bibnamefont {Rubinstein}},\ }\href {https://doi.org/10.1038/nmat2292}
  {\bibfield  {journal} {\bibinfo  {journal} {Nature materials}\ }\textbf
  {\bibinfo {volume} {7}},\ \bibinfo {pages} {997} (\bibinfo {year}
  {2008})}\BibitemShut {NoStop}%
\bibitem [{\citenamefont {Rosa}\ and\ \citenamefont
  {Everaers}(2014)}]{RosaPRL2014}%
  \BibitemOpen
  \bibfield  {author} {\bibinfo {author} {\bibfnamefont {A.}~\bibnamefont
  {Rosa}}\ and\ \bibinfo {author} {\bibfnamefont {R.}~\bibnamefont
  {Everaers}},\ }\href {https://doi.org/10.1103/PhysRevLett.112.118302}
  {\bibfield  {journal} {\bibinfo  {journal} {Phys. Rev. Lett.}\ }\textbf
  {\bibinfo {volume} {112}},\ \bibinfo {pages} {118302} (\bibinfo {year}
  {2014})}\BibitemShut {NoStop}%
\bibitem [{\citenamefont {Michieletto}\ and\ \citenamefont
  {Turner}(2016)}]{MichielettoTurner2016}%
  \BibitemOpen
  \bibfield  {author} {\bibinfo {author} {\bibfnamefont {D.}~\bibnamefont
  {Michieletto}}\ and\ \bibinfo {author} {\bibfnamefont {M.~S.}\ \bibnamefont
  {Turner}},\ }\href@noop {} {\bibfield  {journal} {\bibinfo  {journal} {Proc.
  Natl. Acad. Sci. USA}\ }\textbf {\bibinfo {volume} {113}},\ \bibinfo {pages}
  {5195} (\bibinfo {year} {2016})}\BibitemShut {NoStop}%
\bibitem [{\citenamefont {Ubertini}\ and\ \citenamefont
  {Rosa}(2021)}]{Ubertini_PRE2021}%
  \BibitemOpen
  \bibfield  {author} {\bibinfo {author} {\bibfnamefont {M.~A.}\ \bibnamefont
  {Ubertini}}\ and\ \bibinfo {author} {\bibfnamefont {A.}~\bibnamefont
  {Rosa}},\ }\href@noop {} {\bibfield  {journal} {\bibinfo  {journal} {Phys.
  Rev. E}\ }\textbf {\bibinfo {volume} {104}},\ \bibinfo {pages} {054503}
  (\bibinfo {year} {2021})}\BibitemShut {NoStop}%
\bibitem [{\citenamefont {Ubertini}\ and\ \citenamefont
  {Rosa}(2023)}]{Ubertini_Macromolecules2023}%
  \BibitemOpen
  \bibfield  {author} {\bibinfo {author} {\bibfnamefont {M.~A.}\ \bibnamefont
  {Ubertini}}\ and\ \bibinfo {author} {\bibfnamefont {A.}~\bibnamefont
  {Rosa}},\ }\href {https://doi.org/10.1021/acs.macromol.3c00278} {\bibfield
  {journal} {\bibinfo  {journal} {Macromolecules}\ }\textbf {\bibinfo {volume}
  {56}},\ \bibinfo {pages} {3354} (\bibinfo {year} {2023})}\BibitemShut
  {NoStop}%
\bibitem [{\citenamefont {Tubiana}\ \emph {et~al.}(2024)\citenamefont
  {Tubiana}, \citenamefont {Alexander}, \citenamefont {Barbensi}, \citenamefont
  {Buck}, \citenamefont {Cartwright}, \citenamefont {Chwastyk}, \citenamefont
  {Cieplak}, \citenamefont {Coluzza}, \citenamefont {{\v C}opar}, \citenamefont
  {Craik}, \citenamefont {Di~Stefano}, \citenamefont {Everaers}, \citenamefont
  {Fa{\'\i}sca}, \citenamefont {Ferrari}, \citenamefont {Giacometti},
  \citenamefont {Goundaroulis}, \citenamefont {Haglund}, \citenamefont {Hou},
  \citenamefont {Ilieva}, \citenamefont {Jackson}, \citenamefont {Japaridze},
  \citenamefont {Kaplan}, \citenamefont {Klotz}, \citenamefont {Li},
  \citenamefont {Likos}, \citenamefont {Locatelli}, \citenamefont
  {L{\'o}pez-Le{\'o}n}, \citenamefont {Machon}, \citenamefont {Micheletti},
  \citenamefont {Michieletto}, \citenamefont {Niemi}, \citenamefont {Niemyska},
  \citenamefont {Niewieczerzal}, \citenamefont {Nitti}, \citenamefont
  {Orlandini}, \citenamefont {Pasquali}, \citenamefont {Perlinska},
  \citenamefont {Podgornik}, \citenamefont {Potestio}, \citenamefont {Pugno},
  \citenamefont {Ravnik}, \citenamefont {Ricca}, \citenamefont {Rohwer},
  \citenamefont {Rosa}, \citenamefont {Smrek}, \citenamefont {Souslov},
  \citenamefont {Stasiak}, \citenamefont {Steer}, \citenamefont {Su{\l}kowska},
  \citenamefont {Su{\l}kowski}, \citenamefont {Sumners}, \citenamefont
  {Svaneborg}, \citenamefont {Szymczak}, \citenamefont {Tarenzi}, \citenamefont
  {Travasso}, \citenamefont {Virnau}, \citenamefont {Vlassopoulos},
  \citenamefont {Ziherl},\ and\ \citenamefont {{\v
  Z}umer}}]{TubianaPhysRep2024}%
  \BibitemOpen
  \bibfield  {author} {\bibinfo {author} {\bibfnamefont {L.}~\bibnamefont
  {Tubiana}}, \bibinfo {author} {\bibfnamefont {G.~P.}\ \bibnamefont
  {Alexander}}, \bibinfo {author} {\bibfnamefont {A.}~\bibnamefont {Barbensi}},
  \bibinfo {author} {\bibfnamefont {D.}~\bibnamefont {Buck}}, \bibinfo {author}
  {\bibfnamefont {J.~H.~E.}\ \bibnamefont {Cartwright}}, \bibinfo {author}
  {\bibfnamefont {M.}~\bibnamefont {Chwastyk}}, \bibinfo {author}
  {\bibfnamefont {M.}~\bibnamefont {Cieplak}}, \bibinfo {author} {\bibfnamefont
  {I.}~\bibnamefont {Coluzza}}, \bibinfo {author} {\bibfnamefont
  {S.}~\bibnamefont {{\v C}opar}}, \bibinfo {author} {\bibfnamefont {D.~J.}\
  \bibnamefont {Craik}}, \bibinfo {author} {\bibfnamefont {M.}~\bibnamefont
  {Di~Stefano}}, \bibinfo {author} {\bibfnamefont {R.}~\bibnamefont
  {Everaers}}, \bibinfo {author} {\bibfnamefont {P.~F.~N.}\ \bibnamefont
  {Fa{\'\i}sca}}, \bibinfo {author} {\bibfnamefont {F.}~\bibnamefont
  {Ferrari}}, \bibinfo {author} {\bibfnamefont {A.}~\bibnamefont {Giacometti}},
  \bibinfo {author} {\bibfnamefont {D.}~\bibnamefont {Goundaroulis}}, \bibinfo
  {author} {\bibfnamefont {E.}~\bibnamefont {Haglund}}, \bibinfo {author}
  {\bibfnamefont {Y.-M.}\ \bibnamefont {Hou}}, \bibinfo {author} {\bibfnamefont
  {N.}~\bibnamefont {Ilieva}}, \bibinfo {author} {\bibfnamefont {S.~E.}\
  \bibnamefont {Jackson}}, \bibinfo {author} {\bibfnamefont {A.}~\bibnamefont
  {Japaridze}}, \bibinfo {author} {\bibfnamefont {N.}~\bibnamefont {Kaplan}},
  \bibinfo {author} {\bibfnamefont {A.~R.}\ \bibnamefont {Klotz}}, \bibinfo
  {author} {\bibfnamefont {H.}~\bibnamefont {Li}}, \bibinfo {author}
  {\bibfnamefont {C.~N.}\ \bibnamefont {Likos}}, \bibinfo {author}
  {\bibfnamefont {E.}~\bibnamefont {Locatelli}}, \bibinfo {author}
  {\bibfnamefont {T.}~\bibnamefont {L{\'o}pez-Le{\'o}n}}, \bibinfo {author}
  {\bibfnamefont {T.}~\bibnamefont {Machon}}, \bibinfo {author} {\bibfnamefont
  {C.}~\bibnamefont {Micheletti}}, \bibinfo {author} {\bibfnamefont
  {D.}~\bibnamefont {Michieletto}}, \bibinfo {author} {\bibfnamefont
  {A.}~\bibnamefont {Niemi}}, \bibinfo {author} {\bibfnamefont
  {W.}~\bibnamefont {Niemyska}}, \bibinfo {author} {\bibfnamefont
  {S.}~\bibnamefont {Niewieczerzal}}, \bibinfo {author} {\bibfnamefont
  {F.}~\bibnamefont {Nitti}}, \bibinfo {author} {\bibfnamefont
  {E.}~\bibnamefont {Orlandini}}, \bibinfo {author} {\bibfnamefont
  {S.}~\bibnamefont {Pasquali}}, \bibinfo {author} {\bibfnamefont {A.~P.}\
  \bibnamefont {Perlinska}}, \bibinfo {author} {\bibfnamefont {R.}~\bibnamefont
  {Podgornik}}, \bibinfo {author} {\bibfnamefont {R.}~\bibnamefont {Potestio}},
  \bibinfo {author} {\bibfnamefont {N.~M.}\ \bibnamefont {Pugno}}, \bibinfo
  {author} {\bibfnamefont {M.}~\bibnamefont {Ravnik}}, \bibinfo {author}
  {\bibfnamefont {R.}~\bibnamefont {Ricca}}, \bibinfo {author} {\bibfnamefont
  {C.~M.}\ \bibnamefont {Rohwer}}, \bibinfo {author} {\bibfnamefont
  {A.}~\bibnamefont {Rosa}}, \bibinfo {author} {\bibfnamefont {J.}~\bibnamefont
  {Smrek}}, \bibinfo {author} {\bibfnamefont {A.}~\bibnamefont {Souslov}},
  \bibinfo {author} {\bibfnamefont {A.}~\bibnamefont {Stasiak}}, \bibinfo
  {author} {\bibfnamefont {D.}~\bibnamefont {Steer}}, \bibinfo {author}
  {\bibfnamefont {J.}~\bibnamefont {Su{\l}kowska}}, \bibinfo {author}
  {\bibfnamefont {P.}~\bibnamefont {Su{\l}kowski}}, \bibinfo {author}
  {\bibfnamefont {D.~W.~L.}\ \bibnamefont {Sumners}}, \bibinfo {author}
  {\bibfnamefont {C.}~\bibnamefont {Svaneborg}}, \bibinfo {author}
  {\bibfnamefont {P.}~\bibnamefont {Szymczak}}, \bibinfo {author}
  {\bibfnamefont {T.}~\bibnamefont {Tarenzi}}, \bibinfo {author} {\bibfnamefont
  {R.}~\bibnamefont {Travasso}}, \bibinfo {author} {\bibfnamefont
  {P.}~\bibnamefont {Virnau}}, \bibinfo {author} {\bibfnamefont
  {D.}~\bibnamefont {Vlassopoulos}}, \bibinfo {author} {\bibfnamefont
  {P.}~\bibnamefont {Ziherl}},\ and\ \bibinfo {author} {\bibfnamefont
  {S.}~\bibnamefont {{\v Z}umer}},\ }\href
  {https://doi.org/https://doi.org/10.1016/j.physrep.2024.04.002} {\bibfield
  {journal} {\bibinfo  {journal} {Physics Reports}\ }\textbf {\bibinfo {volume}
  {1075}},\ \bibinfo {pages} {1} (\bibinfo {year} {2024})}\BibitemShut
  {NoStop}%
\bibitem [{\citenamefont {Schroeder}\ \emph {et~al.}(2026)\citenamefont
  {Schroeder}, \citenamefont {Everaers}, \citenamefont {Kremer}, \citenamefont
  {Kruteva}, \citenamefont {Likos}, \citenamefont {McKenna}, \citenamefont
  {O'Connor}, \citenamefont {Ravi~Prakash}, \citenamefont {Richter},
  \citenamefont {Robertson-Anderson}, \citenamefont {Rubinstein}, \citenamefont
  {Schweizer},\ and\ \citenamefont {Vlassopoulos}}]{Schroeder2025}%
  \BibitemOpen
  \bibfield  {author} {\bibinfo {author} {\bibfnamefont {C.~M.}\ \bibnamefont
  {Schroeder}}, \bibinfo {author} {\bibfnamefont {R.}~\bibnamefont {Everaers}},
  \bibinfo {author} {\bibfnamefont {K.}~\bibnamefont {Kremer}}, \bibinfo
  {author} {\bibfnamefont {M.}~\bibnamefont {Kruteva}}, \bibinfo {author}
  {\bibfnamefont {C.~N.}\ \bibnamefont {Likos}}, \bibinfo {author}
  {\bibfnamefont {G.~B.}\ \bibnamefont {McKenna}}, \bibinfo {author}
  {\bibfnamefont {T.}~\bibnamefont {O'Connor}}, \bibinfo {author}
  {\bibfnamefont {J.}~\bibnamefont {Ravi~Prakash}}, \bibinfo {author}
  {\bibfnamefont {D.}~\bibnamefont {Richter}}, \bibinfo {author} {\bibfnamefont
  {R.}~\bibnamefont {Robertson-Anderson}}, \bibinfo {author} {\bibfnamefont
  {M.}~\bibnamefont {Rubinstein}}, \bibinfo {author} {\bibfnamefont {K.~S.}\
  \bibnamefont {Schweizer}},\ and\ \bibinfo {author} {\bibfnamefont
  {D.}~\bibnamefont {Vlassopoulos}},\ }\href
  {https://doi.org/10.1122/8.0001099} {\bibfield  {journal} {\bibinfo
  {journal} {Journal of Rheology}\ }\textbf {\bibinfo {volume} {70}},\ \bibinfo
  {pages} {183} (\bibinfo {year} {2026})}\BibitemShut {NoStop}%
\bibitem [{\citenamefont {Kozik}\ \emph {et~al.}(2019)\citenamefont {Kozik},
  \citenamefont {Rowan}, \citenamefont {Lavelle}, \citenamefont {Berke},
  \citenamefont {Schranz}, \citenamefont {Michelmore},\ and\ \citenamefont
  {Christensen}}]{Kozik2019}%
  \BibitemOpen
  \bibfield  {author} {\bibinfo {author} {\bibfnamefont {A.}~\bibnamefont
  {Kozik}}, \bibinfo {author} {\bibfnamefont {B.~A.}\ \bibnamefont {Rowan}},
  \bibinfo {author} {\bibfnamefont {D.}~\bibnamefont {Lavelle}}, \bibinfo
  {author} {\bibfnamefont {L.}~\bibnamefont {Berke}}, \bibinfo {author}
  {\bibfnamefont {M.~E.}\ \bibnamefont {Schranz}}, \bibinfo {author}
  {\bibfnamefont {R.~W.}\ \bibnamefont {Michelmore}},\ and\ \bibinfo {author}
  {\bibfnamefont {A.~C.}\ \bibnamefont {Christensen}},\ }\href
  {https://doi.org/10.1371/journal.pgen.1008373} {\bibfield  {journal}
  {\bibinfo  {journal} {PLOS Genetics}\ }\textbf {\bibinfo {volume} {15}},\
  \bibinfo {pages} {1} (\bibinfo {year} {2019})}\BibitemShut {NoStop}%
\bibitem [{\citenamefont {Segura}\ \emph {et~al.}(2024)\citenamefont {Segura},
  \citenamefont {D{\'\i}az-Ingelmo}, \citenamefont {Mart{\'\i}nez-Garc{\'\i}a},
  \citenamefont {Ayats-Fraile}, \citenamefont {Nikolaou},\ and\ \citenamefont
  {Roca}}]{Roca2024}%
  \BibitemOpen
  \bibfield  {author} {\bibinfo {author} {\bibfnamefont {J.}~\bibnamefont
  {Segura}}, \bibinfo {author} {\bibfnamefont {O.}~\bibnamefont
  {D{\'\i}az-Ingelmo}}, \bibinfo {author} {\bibfnamefont {B.}~\bibnamefont
  {Mart{\'\i}nez-Garc{\'\i}a}}, \bibinfo {author} {\bibfnamefont
  {A.}~\bibnamefont {Ayats-Fraile}}, \bibinfo {author} {\bibfnamefont
  {C.}~\bibnamefont {Nikolaou}},\ and\ \bibinfo {author} {\bibfnamefont
  {J.}~\bibnamefont {Roca}},\ }\href
  {https://doi.org/10.1038/s41467-024-49023-4} {\bibfield  {journal} {\bibinfo
  {journal} {Nature Communications}\ }\textbf {\bibinfo {volume} {15}},\
  \bibinfo {pages} {4526} (\bibinfo {year} {2024})}\BibitemShut {NoStop}%
\bibitem [{\citenamefont {Jun}\ and\ \citenamefont
  {Mulder}(2006)}]{jun2006entropy}%
  \BibitemOpen
  \bibfield  {author} {\bibinfo {author} {\bibfnamefont {S.}~\bibnamefont
  {Jun}}\ and\ \bibinfo {author} {\bibfnamefont {B.}~\bibnamefont {Mulder}},\
  }\href {https://doi.org/10.1073/pnas.0605305103} {\bibfield  {journal}
  {\bibinfo  {journal} {Proc. Natl. Acad. Sci. USA}\ }\textbf {\bibinfo
  {volume} {103}},\ \bibinfo {pages} {12388} (\bibinfo {year}
  {2006})}\BibitemShut {NoStop}%
\bibitem [{\citenamefont {Junier}\ \emph {et~al.}(2023)\citenamefont {Junier},
  \citenamefont {Ghobadpour}, \citenamefont {Espeli},\ and\ \citenamefont
  {Everaers}}]{Junier2023}%
  \BibitemOpen
  \bibfield  {author} {\bibinfo {author} {\bibfnamefont {I.}~\bibnamefont
  {Junier}}, \bibinfo {author} {\bibfnamefont {E.}~\bibnamefont {Ghobadpour}},
  \bibinfo {author} {\bibfnamefont {O.}~\bibnamefont {Espeli}},\ and\ \bibinfo
  {author} {\bibfnamefont {R.}~\bibnamefont {Everaers}},\ }\href
  {https://www.frontiersin.org/journals/microbiology/articles/10.3389/fmicb.2023.1192831}
  {\bibfield  {journal} {\bibinfo  {journal} {Frontiers in Microbiology}\
  }\textbf {\bibinfo {volume} {14}} (\bibinfo {year} {2023})}\BibitemShut
  {NoStop}%
\bibitem [{\citenamefont {De~Gennes}(1979)}]{de1979scaling}%
  \BibitemOpen
  \bibfield  {author} {\bibinfo {author} {\bibfnamefont {P.-G.}\ \bibnamefont
  {De~Gennes}},\ }\href@noop {} {\emph {\bibinfo {title} {Scaling concepts in
  polymer physics}}}\ (\bibinfo  {publisher} {Cornell university press},\
  \bibinfo {year} {1979})\BibitemShut {NoStop}%
\bibitem [{\citenamefont {Williams}\ \emph {et~al.}(1981)\citenamefont
  {Williams}, \citenamefont {Brochard},\ and\ \citenamefont
  {Frisch}}]{williams1981polymer}%
  \BibitemOpen
  \bibfield  {author} {\bibinfo {author} {\bibfnamefont {C.}~\bibnamefont
  {Williams}}, \bibinfo {author} {\bibfnamefont {F.}~\bibnamefont {Brochard}},\
  and\ \bibinfo {author} {\bibfnamefont {H.}~\bibnamefont {Frisch}},\ }\href
  {https://doi.org/10.1146/annurev.pc.32.100181.002245} {\bibfield  {journal}
  {\bibinfo  {journal} {Annu. Rev. Phys. Chem.}\ }\textbf {\bibinfo {volume}
  {32}},\ \bibinfo {pages} {433} (\bibinfo {year} {1981})}\BibitemShut
  {NoStop}%
\bibitem [{\citenamefont {Khokhlov}\ \emph {et~al.}(1994)\citenamefont
  {Khokhlov}, \citenamefont {Grosberg},\ and\ \citenamefont
  {Pande}}]{khokhlov1994statistical}%
  \BibitemOpen
  \bibfield  {author} {\bibinfo {author} {\bibfnamefont {A.~R.}\ \bibnamefont
  {Khokhlov}}, \bibinfo {author} {\bibfnamefont {A.~Y.}\ \bibnamefont
  {Grosberg}},\ and\ \bibinfo {author} {\bibfnamefont {V.~S.}\ \bibnamefont
  {Pande}},\ }\href@noop {} {\emph {\bibinfo {title} {Statistical physics of
  macromolecules}}},\ Vol.~\bibinfo {volume} {1}\ (\bibinfo  {publisher}
  {Springer},\ \bibinfo {year} {1994})\BibitemShut {NoStop}%
\bibitem [{\citenamefont {Rubinstein}\ and\ \citenamefont
  {Colby}(2003)}]{rubinstein2003polymer}%
  \BibitemOpen
  \bibfield  {author} {\bibinfo {author} {\bibfnamefont {M.}~\bibnamefont
  {Rubinstein}}\ and\ \bibinfo {author} {\bibfnamefont {R.~H.}\ \bibnamefont
  {Colby}},\ }\href {https://doi.org//10.1093/oso/9780198520597.001.0001}
  {\emph {\bibinfo {title} {Polymer physics}}}\ (\bibinfo  {publisher} {Oxford
  university press},\ \bibinfo {year} {2003})\BibitemShut {NoStop}%
\bibitem [{\citenamefont {Rouse~Jr}(1953)}]{rouse1953theory}%
  \BibitemOpen
  \bibfield  {author} {\bibinfo {author} {\bibfnamefont {P.~E.}\ \bibnamefont
  {Rouse~Jr}},\ }\href {https://doi.org/10.1063/1.1699180} {\bibfield
  {journal} {\bibinfo  {journal} {J. Chem. Phys.}\ }\textbf {\bibinfo {volume}
  {21}},\ \bibinfo {pages} {1272} (\bibinfo {year} {1953})}\BibitemShut
  {NoStop}%
\bibitem [{\citenamefont {Zimm}(1956)}]{zimm1956dynamics}%
  \BibitemOpen
  \bibfield  {author} {\bibinfo {author} {\bibfnamefont {B.~H.}\ \bibnamefont
  {Zimm}},\ }\href {https://doi.org//10.1063/1.1742462} {\bibfield  {journal}
  {\bibinfo  {journal} {J. Chem. Phys.}\ }\textbf {\bibinfo {volume} {24}},\
  \bibinfo {pages} {269} (\bibinfo {year} {1956})}\BibitemShut {NoStop}%
\bibitem [{\citenamefont {Smith}\ \emph {et~al.}(1996)\citenamefont {Smith},
  \citenamefont {Perkins},\ and\ \citenamefont {Chu}}]{smith1996dynamical}%
  \BibitemOpen
  \bibfield  {author} {\bibinfo {author} {\bibfnamefont {D.~E.}\ \bibnamefont
  {Smith}}, \bibinfo {author} {\bibfnamefont {T.~T.}\ \bibnamefont {Perkins}},\
  and\ \bibinfo {author} {\bibfnamefont {S.}~\bibnamefont {Chu}},\ }\href
  {https://doi.org/10.1021/ma951455p} {\bibfield  {journal} {\bibinfo
  {journal} {Macromolecules}\ }\textbf {\bibinfo {volume} {29}},\ \bibinfo
  {pages} {1372} (\bibinfo {year} {1996})}\BibitemShut {NoStop}%
\bibitem [{\citenamefont {Hur}\ \emph {et~al.}(2002)\citenamefont {Hur},
  \citenamefont {Shaqfeh}, \citenamefont {Babcock},\ and\ \citenamefont
  {Chu}}]{hur2002dynamics}%
  \BibitemOpen
  \bibfield  {author} {\bibinfo {author} {\bibfnamefont {J.~S.}\ \bibnamefont
  {Hur}}, \bibinfo {author} {\bibfnamefont {E.~S.}\ \bibnamefont {Shaqfeh}},
  \bibinfo {author} {\bibfnamefont {H.~P.}\ \bibnamefont {Babcock}},\ and\
  \bibinfo {author} {\bibfnamefont {S.}~\bibnamefont {Chu}},\ }\href
  {https://doi.org/10.1103/PhysRevE.66.011915} {\bibfield  {journal} {\bibinfo
  {journal} {Phys. Rev. E}\ }\textbf {\bibinfo {volume} {66}},\ \bibinfo
  {pages} {011915} (\bibinfo {year} {2002})}\BibitemShut {NoStop}%
\bibitem [{\citenamefont {Hur}\ \emph {et~al.}(2000)\citenamefont {Hur},
  \citenamefont {Shaqfeh},\ and\ \citenamefont {Larson}}]{hur2000brownian}%
  \BibitemOpen
  \bibfield  {author} {\bibinfo {author} {\bibfnamefont {J.~S.}\ \bibnamefont
  {Hur}}, \bibinfo {author} {\bibfnamefont {E.~S.}\ \bibnamefont {Shaqfeh}},\
  and\ \bibinfo {author} {\bibfnamefont {R.~G.}\ \bibnamefont {Larson}},\
  }\href {https://doi.org/10.1122/1.551115} {\bibfield  {journal} {\bibinfo
  {journal} {J. Rheol.}\ }\textbf {\bibinfo {volume} {44}},\ \bibinfo {pages}
  {713} (\bibinfo {year} {2000})}\BibitemShut {NoStop}%
\bibitem [{\citenamefont {Smith}\ \emph {et~al.}(1999)\citenamefont {Smith},
  \citenamefont {Babcock},\ and\ \citenamefont {Chu}}]{smith1999single}%
  \BibitemOpen
  \bibfield  {author} {\bibinfo {author} {\bibfnamefont {D.~E.}\ \bibnamefont
  {Smith}}, \bibinfo {author} {\bibfnamefont {H.~P.}\ \bibnamefont {Babcock}},\
  and\ \bibinfo {author} {\bibfnamefont {S.}~\bibnamefont {Chu}},\ }\href
  {https://doi.org/10.1126/science.283.5408.1724} {\bibfield  {journal}
  {\bibinfo  {journal} {Science}\ }\textbf {\bibinfo {volume} {283}},\ \bibinfo
  {pages} {1724} (\bibinfo {year} {1999})}\BibitemShut {NoStop}%
\bibitem [{\citenamefont {Kaznessis}\ \emph {et~al.}(1999)\citenamefont
  {Kaznessis}, \citenamefont {Hill},\ and\ \citenamefont
  {Maginn}}]{kaznessis1999dielectric}%
  \BibitemOpen
  \bibfield  {author} {\bibinfo {author} {\bibfnamefont {Y.~N.}\ \bibnamefont
  {Kaznessis}}, \bibinfo {author} {\bibfnamefont {D.~A.}\ \bibnamefont
  {Hill}},\ and\ \bibinfo {author} {\bibfnamefont {E.~J.}\ \bibnamefont
  {Maginn}},\ }\href {https://doi.org/10.1021/ma990680u} {\bibfield  {journal}
  {\bibinfo  {journal} {Macromolecules}\ }\textbf {\bibinfo {volume} {32}},\
  \bibinfo {pages} {6679} (\bibinfo {year} {1999})}\BibitemShut {NoStop}%
\bibitem [{\citenamefont {Dunweg}\ and\ \citenamefont
  {Kremer}(1991)}]{dunweg1991microscopic}%
  \BibitemOpen
  \bibfield  {author} {\bibinfo {author} {\bibfnamefont {B.}~\bibnamefont
  {Dunweg}}\ and\ \bibinfo {author} {\bibfnamefont {K.}~\bibnamefont
  {Kremer}},\ }\href {https://doi.org//10.1103/PhysRevLett.66.2996} {\bibfield
  {journal} {\bibinfo  {journal} {Phys. Rev. Lett.}\ }\textbf {\bibinfo
  {volume} {66}},\ \bibinfo {pages} {2996} (\bibinfo {year}
  {1991})}\BibitemShut {NoStop}%
\bibitem [{\citenamefont {Pierleoni}\ and\ \citenamefont
  {Ryckaert}(1991)}]{pierleoni1991relaxation}%
  \BibitemOpen
  \bibfield  {author} {\bibinfo {author} {\bibfnamefont {C.}~\bibnamefont
  {Pierleoni}}\ and\ \bibinfo {author} {\bibfnamefont {J.-P.}\ \bibnamefont
  {Ryckaert}},\ }\href {https://doi.org/10.1103/PhysRevLett.66.2992} {\bibfield
   {journal} {\bibinfo  {journal} {Phys. Rev. Lett.}\ }\textbf {\bibinfo
  {volume} {66}},\ \bibinfo {pages} {2992} (\bibinfo {year}
  {1991})}\BibitemShut {NoStop}%
\bibitem [{\citenamefont {D{\"u}nweg}\ and\ \citenamefont
  {Kremer}(1993)}]{dunweg1993molecular}%
  \BibitemOpen
  \bibfield  {author} {\bibinfo {author} {\bibfnamefont {B.}~\bibnamefont
  {D{\"u}nweg}}\ and\ \bibinfo {author} {\bibfnamefont {K.}~\bibnamefont
  {Kremer}},\ }\href {https://doi.org//10.1063/1.465445} {\bibfield  {journal}
  {\bibinfo  {journal} {J. Chem. Phys.}\ }\textbf {\bibinfo {volume} {99}},\
  \bibinfo {pages} {6983} (\bibinfo {year} {1993})}\BibitemShut {NoStop}%
\bibitem [{\citenamefont {Polson}\ and\ \citenamefont
  {Gallant}(2006)}]{polson2006equilibrium}%
  \BibitemOpen
  \bibfield  {author} {\bibinfo {author} {\bibfnamefont {J.~M.}\ \bibnamefont
  {Polson}}\ and\ \bibinfo {author} {\bibfnamefont {J.~P.}\ \bibnamefont
  {Gallant}},\ }\bibfield  {journal} {\bibinfo  {journal} {J. Chem. Phys.}\
  }\textbf {\bibinfo {volume} {124}},\ \href
  {https://doi.org//10.1063/1.2194903} {/10.1063/1.2194903} (\bibinfo {year}
  {2006})\BibitemShut {NoStop}%
\bibitem [{\citenamefont {Hoogerbrugge}\ and\ \citenamefont
  {Koelman}(1992)}]{hoogerbrugge1992simulating}%
  \BibitemOpen
  \bibfield  {author} {\bibinfo {author} {\bibfnamefont {P.}~\bibnamefont
  {Hoogerbrugge}}\ and\ \bibinfo {author} {\bibfnamefont {J.}~\bibnamefont
  {Koelman}},\ }\href {https://doi.org/10.1209/0295-5075/19/3/001} {\bibfield
  {journal} {\bibinfo  {journal} {EPL}\ }\textbf {\bibinfo {volume} {19}},\
  \bibinfo {pages} {155} (\bibinfo {year} {1992})}\BibitemShut {NoStop}%
\bibitem [{\citenamefont {Espanol}\ and\ \citenamefont
  {Warren}(1995)}]{espanol1995statistical}%
  \BibitemOpen
  \bibfield  {author} {\bibinfo {author} {\bibfnamefont {P.}~\bibnamefont
  {Espanol}}\ and\ \bibinfo {author} {\bibfnamefont {P.}~\bibnamefont
  {Warren}},\ }\href {https://doi.org/10.1209/0295-5075/30/4/001} {\bibfield
  {journal} {\bibinfo  {journal} {EPL}\ }\textbf {\bibinfo {volume} {30}},\
  \bibinfo {pages} {191} (\bibinfo {year} {1995})}\BibitemShut {NoStop}%
\bibitem [{\citenamefont {Ripoll}\ \emph {et~al.}(2001)\citenamefont {Ripoll},
  \citenamefont {Ernst},\ and\ \citenamefont {Espanol}}]{ripoll2001large}%
  \BibitemOpen
  \bibfield  {author} {\bibinfo {author} {\bibfnamefont {M.}~\bibnamefont
  {Ripoll}}, \bibinfo {author} {\bibfnamefont {M.}~\bibnamefont {Ernst}},\ and\
  \bibinfo {author} {\bibfnamefont {P.}~\bibnamefont {Espanol}},\ }\href
  {https://doi.org/10.1063/1.1402989} {\bibfield  {journal} {\bibinfo
  {journal} {J. Chem. Phys.}\ }\textbf {\bibinfo {volume} {115}},\ \bibinfo
  {pages} {7271} (\bibinfo {year} {2001})}\BibitemShut {NoStop}%
\bibitem [{\citenamefont {Groot}\ and\ \citenamefont
  {Warren}(1997)}]{groot1997dissipative}%
  \BibitemOpen
  \bibfield  {author} {\bibinfo {author} {\bibfnamefont {R.~D.}\ \bibnamefont
  {Groot}}\ and\ \bibinfo {author} {\bibfnamefont {P.~B.}\ \bibnamefont
  {Warren}},\ }\href {https://doi.org/10.1063/1.474784} {\bibfield  {journal}
  {\bibinfo  {journal} {J. Chem. Phys.}\ }\textbf {\bibinfo {volume} {107}},\
  \bibinfo {pages} {4423} (\bibinfo {year} {1997})}\BibitemShut {NoStop}%
\bibitem [{\citenamefont {Espanol}\ and\ \citenamefont
  {Warren}(2017)}]{espanol2017perspective}%
  \BibitemOpen
  \bibfield  {author} {\bibinfo {author} {\bibfnamefont {P.}~\bibnamefont
  {Espanol}}\ and\ \bibinfo {author} {\bibfnamefont {P.~B.}\ \bibnamefont
  {Warren}},\ }\href {https://doi.org/10.1063/1.4979514} {\bibfield  {journal}
  {\bibinfo  {journal} {J. Chem Phys.}\ }\textbf {\bibinfo {volume} {146}},\
  \bibinfo {pages} {150901} (\bibinfo {year} {2017})}\BibitemShut {NoStop}%
\bibitem [{\citenamefont {Singh}\ \emph {et~al.}(2021)\citenamefont {Singh},
  \citenamefont {Chauhan}, \citenamefont {Puri},\ and\ \citenamefont
  {Singh}}]{singh2021photo}%
  \BibitemOpen
  \bibfield  {author} {\bibinfo {author} {\bibfnamefont {A.~K.}\ \bibnamefont
  {Singh}}, \bibinfo {author} {\bibfnamefont {A.}~\bibnamefont {Chauhan}},
  \bibinfo {author} {\bibfnamefont {S.}~\bibnamefont {Puri}},\ and\ \bibinfo
  {author} {\bibfnamefont {A.}~\bibnamefont {Singh}},\ }\href
  {https://doi.org/10.1039/D0SM01664K} {\bibfield  {journal} {\bibinfo
  {journal} {Soft Matter}\ }\textbf {\bibinfo {volume} {17}},\ \bibinfo {pages}
  {1802} (\bibinfo {year} {2021})}\BibitemShut {NoStop}%
\bibitem [{\citenamefont {Groot}\ \emph {et~al.}(1999)\citenamefont {Groot},
  \citenamefont {Madden},\ and\ \citenamefont {Tildesley}}]{groot1999role}%
  \BibitemOpen
  \bibfield  {author} {\bibinfo {author} {\bibfnamefont {R.~D.}\ \bibnamefont
  {Groot}}, \bibinfo {author} {\bibfnamefont {T.~J.}\ \bibnamefont {Madden}},\
  and\ \bibinfo {author} {\bibfnamefont {D.~J.}\ \bibnamefont {Tildesley}},\
  }\href {https://doi.org/10.1063/1.478939} {\bibfield  {journal} {\bibinfo
  {journal} {J. Chem. Phys.}\ }\textbf {\bibinfo {volume} {110}},\ \bibinfo
  {pages} {9739} (\bibinfo {year} {1999})}\BibitemShut {NoStop}%
\bibitem [{\citenamefont {Laradji}\ and\ \citenamefont
  {Hore}(2004)}]{laradji2004nanospheres}%
  \BibitemOpen
  \bibfield  {author} {\bibinfo {author} {\bibfnamefont {M.}~\bibnamefont
  {Laradji}}\ and\ \bibinfo {author} {\bibfnamefont {M.~J.}\ \bibnamefont
  {Hore}},\ }\href {https://doi.org/10.1063/1.1806815} {\bibfield  {journal}
  {\bibinfo  {journal} {J. Chem. Phys.}\ }\textbf {\bibinfo {volume} {121}},\
  \bibinfo {pages} {10641} (\bibinfo {year} {2004})}\BibitemShut {NoStop}%
\bibitem [{\citenamefont {Laradji}\ and\ \citenamefont
  {Sunil~Kumar}(2004)}]{laradji2004dynamics}%
  \BibitemOpen
  \bibfield  {author} {\bibinfo {author} {\bibfnamefont {M.}~\bibnamefont
  {Laradji}}\ and\ \bibinfo {author} {\bibfnamefont {P.}~\bibnamefont
  {Sunil~Kumar}},\ }\href {https://doi.org/10.1103/PhysRevLett.93.198105}
  {\bibfield  {journal} {\bibinfo  {journal} {Phys. Rev. Lett.}\ }\textbf
  {\bibinfo {volume} {93}},\ \bibinfo {pages} {198105} (\bibinfo {year}
  {2004})}\BibitemShut {NoStop}%
\bibitem [{\citenamefont {Wijmans}\ and\ \citenamefont
  {Smit}(2002)}]{wijmans2002simulating}%
  \BibitemOpen
  \bibfield  {author} {\bibinfo {author} {\bibfnamefont {C.}~\bibnamefont
  {Wijmans}}\ and\ \bibinfo {author} {\bibfnamefont {B.}~\bibnamefont {Smit}},\
  }\href {https://doi.org/10.1021/ma020086b} {\bibfield  {journal} {\bibinfo
  {journal} {Macromolecules}\ }\textbf {\bibinfo {volume} {35}},\ \bibinfo
  {pages} {7138} (\bibinfo {year} {2002})}\BibitemShut {NoStop}%
\bibitem [{\citenamefont {Huang}\ \emph {et~al.}(2006)\citenamefont {Huang},
  \citenamefont {Wang},\ and\ \citenamefont {Laradji}}]{huang2006flow}%
  \BibitemOpen
  \bibfield  {author} {\bibinfo {author} {\bibfnamefont {J.}~\bibnamefont
  {Huang}}, \bibinfo {author} {\bibfnamefont {Y.}~\bibnamefont {Wang}},\ and\
  \bibinfo {author} {\bibfnamefont {M.}~\bibnamefont {Laradji}},\ }\href
  {https://doi.org/10.1021/ma060628f} {\bibfield  {journal} {\bibinfo
  {journal} {Macromolecules}\ }\textbf {\bibinfo {volume} {39}},\ \bibinfo
  {pages} {5546} (\bibinfo {year} {2006})}\BibitemShut {NoStop}%
\bibitem [{\citenamefont {Groot}(2006)}]{groot2006local}%
  \BibitemOpen
  \bibfield  {author} {\bibinfo {author} {\bibfnamefont {R.~D.}\ \bibnamefont
  {Groot}},\ }\href {https://doi.org/10.1021/ct050269e} {\bibfield  {journal}
  {\bibinfo  {journal} {J. Chem. Theory Comput.}\ }\textbf {\bibinfo {volume}
  {2}},\ \bibinfo {pages} {568} (\bibinfo {year} {2006})}\BibitemShut {NoStop}%
\bibitem [{\citenamefont {Nikunen}\ \emph {et~al.}(2003)\citenamefont
  {Nikunen}, \citenamefont {Karttunen},\ and\ \citenamefont
  {Vattulainen}}]{nikunen2003would}%
  \BibitemOpen
  \bibfield  {author} {\bibinfo {author} {\bibfnamefont {P.}~\bibnamefont
  {Nikunen}}, \bibinfo {author} {\bibfnamefont {M.}~\bibnamefont {Karttunen}},\
  and\ \bibinfo {author} {\bibfnamefont {I.}~\bibnamefont {Vattulainen}},\
  }\href {https://doi.org/10.1016/S0010-4655(03)00202-9} {\bibfield  {journal}
  {\bibinfo  {journal} {Comput. Phys. Commun.}\ }\textbf {\bibinfo {volume}
  {153}},\ \bibinfo {pages} {407} (\bibinfo {year} {2003})}\BibitemShut
  {NoStop}%
\bibitem [{\citenamefont {Nikunen}\ \emph {et~al.}(2007)\citenamefont
  {Nikunen}, \citenamefont {Vattulainen},\ and\ \citenamefont
  {Karttunen}}]{nikunen2007reptational}%
  \BibitemOpen
  \bibfield  {author} {\bibinfo {author} {\bibfnamefont {P.}~\bibnamefont
  {Nikunen}}, \bibinfo {author} {\bibfnamefont {I.}~\bibnamefont
  {Vattulainen}},\ and\ \bibinfo {author} {\bibfnamefont {M.}~\bibnamefont
  {Karttunen}},\ }\href {https://doi.org/10.1103/PhysRevE.75.036713} {\bibfield
   {journal} {\bibinfo  {journal} {Phys. Rev. E}\ }\textbf {\bibinfo {volume}
  {75}},\ \bibinfo {pages} {036713} (\bibinfo {year} {2007})}\BibitemShut
  {NoStop}%
\bibitem [{\citenamefont {Groot}\ and\ \citenamefont
  {Madden}(1998)}]{groot1998dynamic}%
  \BibitemOpen
  \bibfield  {author} {\bibinfo {author} {\bibfnamefont {R.~D.}\ \bibnamefont
  {Groot}}\ and\ \bibinfo {author} {\bibfnamefont {T.~J.}\ \bibnamefont
  {Madden}},\ }\href {https://doi.org/10.1063/1.476300} {\bibfield  {journal}
  {\bibinfo  {journal} {J. Chem. Phys.}\ }\textbf {\bibinfo {volume} {108}},\
  \bibinfo {pages} {8713} (\bibinfo {year} {1998})}\BibitemShut {NoStop}%
\bibitem [{\citenamefont {Singh}\ \emph {et~al.}(2018)\citenamefont {Singh},
  \citenamefont {Chakraborti},\ and\ \citenamefont {Singh}}]{singh2018role}%
  \BibitemOpen
  \bibfield  {author} {\bibinfo {author} {\bibfnamefont {A.}~\bibnamefont
  {Singh}}, \bibinfo {author} {\bibfnamefont {A.}~\bibnamefont {Chakraborti}},\
  and\ \bibinfo {author} {\bibfnamefont {A.}~\bibnamefont {Singh}},\ }\href
  {https://doi.org/10.1039/C8SM00625C} {\bibfield  {journal} {\bibinfo
  {journal} {Soft Matter}\ }\textbf {\bibinfo {volume} {14}},\ \bibinfo {pages}
  {4317} (\bibinfo {year} {2018})}\BibitemShut {NoStop}%
\bibitem [{\citenamefont {Kremer}\ and\ \citenamefont
  {Grest}(1990{\natexlab{a}})}]{kremer1990dynamics}%
  \BibitemOpen
  \bibfield  {author} {\bibinfo {author} {\bibfnamefont {K.}~\bibnamefont
  {Kremer}}\ and\ \bibinfo {author} {\bibfnamefont {G.~S.}\ \bibnamefont
  {Grest}},\ }\href {https://doi.org/10.1063/1.458541} {\bibfield  {journal}
  {\bibinfo  {journal} {The Journal of Chemical Physics}\ }\textbf {\bibinfo
  {volume} {92}},\ \bibinfo {pages} {5057} (\bibinfo {year}
  {1990}{\natexlab{a}})}\BibitemShut {NoStop}%
\bibitem [{\citenamefont {Junghans}\ \emph {et~al.}(2008)\citenamefont
  {Junghans}, \citenamefont {Praprotnik},\ and\ \citenamefont
  {Kremer}}]{junghans2008transport}%
  \BibitemOpen
  \bibfield  {author} {\bibinfo {author} {\bibfnamefont {C.}~\bibnamefont
  {Junghans}}, \bibinfo {author} {\bibfnamefont {M.}~\bibnamefont
  {Praprotnik}},\ and\ \bibinfo {author} {\bibfnamefont {K.}~\bibnamefont
  {Kremer}},\ }\href {https://doi.org/10.1039/B713568H} {\bibfield  {journal}
  {\bibinfo  {journal} {Soft Matter}\ }\textbf {\bibinfo {volume} {4}},\
  \bibinfo {pages} {156} (\bibinfo {year} {2008})}\BibitemShut {NoStop}%
\bibitem [{\citenamefont {Hsu}\ and\ \citenamefont
  {Kremer}(2016)}]{hsu2016static}%
  \BibitemOpen
  \bibfield  {author} {\bibinfo {author} {\bibfnamefont {H.-P.}\ \bibnamefont
  {Hsu}}\ and\ \bibinfo {author} {\bibfnamefont {K.}~\bibnamefont {Kremer}},\
  }\href {https://doi.org/10.1063/1.4946033} {\bibfield  {journal} {\bibinfo
  {journal} {J. Chem. Phys.}\ }\textbf {\bibinfo {volume} {144}},\ \bibinfo
  {pages} {154907} (\bibinfo {year} {2016})}\BibitemShut {NoStop}%
\bibitem [{\citenamefont {Plimpton}(1995)}]{plimpton1995fast}%
  \BibitemOpen
  \bibfield  {author} {\bibinfo {author} {\bibfnamefont {S.}~\bibnamefont
  {Plimpton}},\ }\href {https://doi.org/10.1006/jcph.1995.1039} {\bibfield
  {journal} {\bibinfo  {journal} {J. Comput. Phys.}\ }\textbf {\bibinfo
  {volume} {117}},\ \bibinfo {pages} {1} (\bibinfo {year} {1995})}\BibitemShut
  {NoStop}%
\bibitem [{\citenamefont {Singh}\ \emph {et~al.}(2016)\citenamefont {Singh},
  \citenamefont {Kuksenok}, \citenamefont {Johnson},\ and\ \citenamefont
  {Balazs}}]{singh2016tailoring}%
  \BibitemOpen
  \bibfield  {author} {\bibinfo {author} {\bibfnamefont {A.}~\bibnamefont
  {Singh}}, \bibinfo {author} {\bibfnamefont {O.}~\bibnamefont {Kuksenok}},
  \bibinfo {author} {\bibfnamefont {J.~A.}\ \bibnamefont {Johnson}},\ and\
  \bibinfo {author} {\bibfnamefont {A.~C.}\ \bibnamefont {Balazs}},\ }\href
  {https://doi.org/10.1039/C6PY00325G} {\bibfield  {journal} {\bibinfo
  {journal} {Polymer Chemistry}\ }\textbf {\bibinfo {volume} {7}},\ \bibinfo
  {pages} {2955} (\bibinfo {year} {2016})}\BibitemShut {NoStop}%
\bibitem [{\citenamefont {Singh}\ \emph {et~al.}(2017)\citenamefont {Singh},
  \citenamefont {Kuksenok}, \citenamefont {Johnson},\ and\ \citenamefont
  {Balazs}}]{singh2017photo}%
  \BibitemOpen
  \bibfield  {author} {\bibinfo {author} {\bibfnamefont {A.}~\bibnamefont
  {Singh}}, \bibinfo {author} {\bibfnamefont {O.}~\bibnamefont {Kuksenok}},
  \bibinfo {author} {\bibfnamefont {J.~A.}\ \bibnamefont {Johnson}},\ and\
  \bibinfo {author} {\bibfnamefont {A.~C.}\ \bibnamefont {Balazs}},\ }\href
  {https://doi.org/10.1039/C6SM02625G} {\bibfield  {journal} {\bibinfo
  {journal} {Soft Matter}\ }\textbf {\bibinfo {volume} {13}},\ \bibinfo {pages}
  {1978} (\bibinfo {year} {2017})}\BibitemShut {NoStop}%
\bibitem [{\citenamefont {Sirk}\ \emph {et~al.}(2012)\citenamefont {Sirk},
  \citenamefont {Slizoberg}, \citenamefont {Brennan}, \citenamefont {Lisal},\
  and\ \citenamefont {Andzelm}}]{sirk2012enhanced}%
  \BibitemOpen
  \bibfield  {author} {\bibinfo {author} {\bibfnamefont {T.~W.}\ \bibnamefont
  {Sirk}}, \bibinfo {author} {\bibfnamefont {Y.~R.}\ \bibnamefont {Slizoberg}},
  \bibinfo {author} {\bibfnamefont {J.~K.}\ \bibnamefont {Brennan}}, \bibinfo
  {author} {\bibfnamefont {M.}~\bibnamefont {Lisal}},\ and\ \bibinfo {author}
  {\bibfnamefont {J.~W.}\ \bibnamefont {Andzelm}},\ }\href
  {https://doi.org/10.1063/1.3698476} {\bibfield  {journal} {\bibinfo
  {journal} {J. Chem. Phys.}\ }\textbf {\bibinfo {volume} {136}},\ \bibinfo
  {pages} {134903} (\bibinfo {year} {2012})}\BibitemShut {NoStop}%
\bibitem [{\citenamefont {Singh}\ and\ \citenamefont
  {Singh}(2023)}]{singh2023phase}%
  \BibitemOpen
  \bibfield  {author} {\bibinfo {author} {\bibfnamefont {A.~K.}\ \bibnamefont
  {Singh}}\ and\ \bibinfo {author} {\bibfnamefont {A.}~\bibnamefont {Singh}},\
  }\href {https://doi.org/10.1016/j.commatsci.2023.112224} {\bibfield
  {journal} {\bibinfo  {journal} {Comput. Mater. Sci.}\ }\textbf {\bibinfo
  {volume} {226}},\ \bibinfo {pages} {112224} (\bibinfo {year}
  {2023})}\BibitemShut {NoStop}%
\bibitem [{\citenamefont {Gogoi}\ \emph {et~al.}(2023)\citenamefont {Gogoi},
  \citenamefont {Chauhan}, \citenamefont {Puri},\ and\ \citenamefont
  {Singh}}]{singh2023additive}%
  \BibitemOpen
  \bibfield  {author} {\bibinfo {author} {\bibfnamefont {D.}~\bibnamefont
  {Gogoi}}, \bibinfo {author} {\bibfnamefont {A.}~\bibnamefont {Chauhan}},
  \bibinfo {author} {\bibfnamefont {S.}~\bibnamefont {Puri}},\ and\ \bibinfo
  {author} {\bibfnamefont {A.}~\bibnamefont {Singh}},\ }\href
  {https://doi.org/10.1039/D3SM00504F} {\bibfield  {journal} {\bibinfo
  {journal} {Soft Matter}\ }\textbf {\bibinfo {volume} {19}},\ \bibinfo {pages}
  {6433} (\bibinfo {year} {2023})}\BibitemShut {NoStop}%
\bibitem [{\citenamefont {Chauhan}\ \emph {et~al.}(2023)\citenamefont
  {Chauhan}, \citenamefont {Gogoi}, \citenamefont {Puri},\ and\ \citenamefont
  {Singh}}]{singh2023amph}%
  \BibitemOpen
  \bibfield  {author} {\bibinfo {author} {\bibfnamefont {A.}~\bibnamefont
  {Chauhan}}, \bibinfo {author} {\bibfnamefont {D.}~\bibnamefont {Gogoi}},
  \bibinfo {author} {\bibfnamefont {S.}~\bibnamefont {Puri}},\ and\ \bibinfo
  {author} {\bibfnamefont {A.}~\bibnamefont {Singh}},\ }\href
  {https://doi.org/10.1063/5.0173817} {\bibfield  {journal} {\bibinfo
  {journal} {The Journal of Chemical Physics}\ }\textbf {\bibinfo {volume}
  {159}},\ \bibinfo {pages} {204901} (\bibinfo {year} {2023})}\BibitemShut
  {NoStop}%
\bibitem [{\citenamefont {Hashimoto}\ \emph {et~al.}(1986)\citenamefont
  {Hashimoto}, \citenamefont {Itakura},\ and\ \citenamefont
  {Hasegawa}}]{hashimoto1986late}%
  \BibitemOpen
  \bibfield  {author} {\bibinfo {author} {\bibfnamefont {T.}~\bibnamefont
  {Hashimoto}}, \bibinfo {author} {\bibfnamefont {M.}~\bibnamefont {Itakura}},\
  and\ \bibinfo {author} {\bibfnamefont {H.}~\bibnamefont {Hasegawa}},\ }\href
  {https://doi.org/10.1063/1.451477} {\bibfield  {journal} {\bibinfo  {journal}
  {J. Chem. Phys.}\ }\textbf {\bibinfo {volume} {85}},\ \bibinfo {pages} {6118}
  (\bibinfo {year} {1986})}\BibitemShut {NoStop}%
\bibitem [{\citenamefont {Shrivastava}\ \emph {et~al.}(2024)\citenamefont
  {Shrivastava}, \citenamefont {Upadhyay}, \citenamefont {Pradhan},
  \citenamefont {Saha},\ and\ \citenamefont {Singh}}]{sam2024emulsion}%
  \BibitemOpen
  \bibfield  {author} {\bibinfo {author} {\bibfnamefont {S.}~\bibnamefont
  {Shrivastava}}, \bibinfo {author} {\bibfnamefont {A.}~\bibnamefont
  {Upadhyay}}, \bibinfo {author} {\bibfnamefont {S.~S.}\ \bibnamefont
  {Pradhan}}, \bibinfo {author} {\bibfnamefont {S.}~\bibnamefont {Saha}},\ and\
  \bibinfo {author} {\bibfnamefont {A.}~\bibnamefont {Singh}},\ }\href
  {https://doi.org/10.1021/acs.langmuir.4c01083} {\bibfield  {journal}
  {\bibinfo  {journal} {Langmuir}\ }\textbf {\bibinfo {volume} {40}},\ \bibinfo
  {pages} {13920} (\bibinfo {year} {2024})}\BibitemShut {NoStop}%
\bibitem [{\citenamefont {Shrivastava}\ \emph {et~al.}(2022)\citenamefont
  {Shrivastava}, \citenamefont {Ifra}, \citenamefont {Saha},\ and\
  \citenamefont {Singh}}]{sam2022brush}%
  \BibitemOpen
  \bibfield  {author} {\bibinfo {author} {\bibfnamefont {S.}~\bibnamefont
  {Shrivastava}}, \bibinfo {author} {\bibnamefont {Ifra}}, \bibinfo {author}
  {\bibfnamefont {S.}~\bibnamefont {Saha}},\ and\ \bibinfo {author}
  {\bibfnamefont {A.}~\bibnamefont {Singh}},\ }\href@noop {} {\bibfield
  {journal} {\bibinfo  {journal} {Phys. Chem. Chem. Phys.}\ }\textbf {\bibinfo
  {volume} {24}},\ \bibinfo {pages} {17986} (\bibinfo {year}
  {2022})}\BibitemShut {NoStop}%
\bibitem [{\citenamefont {Symeonidis}\ \emph {et~al.}(2006)\citenamefont
  {Symeonidis}, \citenamefont {Karniadakis},\ and\ \citenamefont
  {Caswell}}]{symeonidis2006schmidt}%
  \BibitemOpen
  \bibfield  {author} {\bibinfo {author} {\bibfnamefont {V.}~\bibnamefont
  {Symeonidis}}, \bibinfo {author} {\bibfnamefont {G.~E.}\ \bibnamefont
  {Karniadakis}},\ and\ \bibinfo {author} {\bibfnamefont {B.}~\bibnamefont
  {Caswell}},\ }\href {https://doi.org/10.1063/1.2360274} {\bibfield  {journal}
  {\bibinfo  {journal} {J. Chem. Phys.}\ }\textbf {\bibinfo {volume} {125}},\
  \bibinfo {pages} {184902} (\bibinfo {year} {2006})}\BibitemShut {NoStop}%
\bibitem [{\citenamefont {Kremer}\ and\ \citenamefont
  {Grest}(1990{\natexlab{b}})}]{Kremer1990}%
  \BibitemOpen
  \bibfield  {author} {\bibinfo {author} {\bibfnamefont {K.}~\bibnamefont
  {Kremer}}\ and\ \bibinfo {author} {\bibfnamefont {G.~S.}\ \bibnamefont
  {Grest}},\ }\href {https://doi.org/10.1088/0953-8984/2/S/045} {\bibfield
  {journal} {\bibinfo  {journal} {Journal of Physics: Condensed Matter}\
  }\textbf {\bibinfo {volume} {2}},\ \bibinfo {pages} {SA295} (\bibinfo {year}
  {1990}{\natexlab{b}})}\BibitemShut {NoStop}%
\bibitem [{\citenamefont {Liu}\ \emph {et~al.}(2018)\citenamefont {Liu},
  \citenamefont {den Otter},\ and\ \citenamefont {Briels}}]{Wim2018}%
  \BibitemOpen
  \bibfield  {author} {\bibinfo {author} {\bibfnamefont {L.}~\bibnamefont
  {Liu}}, \bibinfo {author} {\bibfnamefont {W.~K.}\ \bibnamefont {den Otter}},\
  and\ \bibinfo {author} {\bibfnamefont {W.~J.}\ \bibnamefont {Briels}},\
  }\href {https://doi.org/10.1021/acs.jpcb.8b03104} {\bibfield  {journal}
  {\bibinfo  {journal} {The Journal of Physical Chemistry B}\ }\textbf
  {\bibinfo {volume} {122}},\ \bibinfo {pages} {10210} (\bibinfo {year}
  {2018})}\BibitemShut {NoStop}%
\bibitem [{\citenamefont {Yong}\ \emph {et~al.}(2013)\citenamefont {Yong},
  \citenamefont {Kuksenok}, \citenamefont {Matyjaszewski},\ and\ \citenamefont
  {Balazs}}]{Yong2013}%
  \BibitemOpen
  \bibfield  {author} {\bibinfo {author} {\bibfnamefont {X.}~\bibnamefont
  {Yong}}, \bibinfo {author} {\bibfnamefont {O.}~\bibnamefont {Kuksenok}},
  \bibinfo {author} {\bibfnamefont {K.}~\bibnamefont {Matyjaszewski}},\ and\
  \bibinfo {author} {\bibfnamefont {A.~C.}\ \bibnamefont {Balazs}},\ }\href
  {https://doi.org/10.1021/nl403855k} {\bibfield  {journal} {\bibinfo
  {journal} {Nano Letters}\ }\textbf {\bibinfo {volume} {13}},\ \bibinfo
  {pages} {6269} (\bibinfo {year} {2013})}\BibitemShut {NoStop}%
\bibitem [{\citenamefont {Nikoobakht}\ and\ \citenamefont
  {El-Sayed}(2003)}]{nikoobakht2003preparation}%
  \BibitemOpen
  \bibfield  {author} {\bibinfo {author} {\bibfnamefont {B.}~\bibnamefont
  {Nikoobakht}}\ and\ \bibinfo {author} {\bibfnamefont {M.~A.}\ \bibnamefont
  {El-Sayed}},\ }\href {https://doi.org/10.1021/cm020732l} {\bibfield
  {journal} {\bibinfo  {journal} {Chem. Mater.}\ }\textbf {\bibinfo {volume}
  {15}},\ \bibinfo {pages} {1957} (\bibinfo {year} {2003})}\BibitemShut
  {NoStop}%
\bibitem [{\citenamefont {Mohorič}\ \emph {et~al.}(2020)\citenamefont
  {Mohorič}, \citenamefont {Lahajnar},\ and\ \citenamefont
  {Stepišnik}}]{Janez2020}%
  \BibitemOpen
  \bibfield  {author} {\bibinfo {author} {\bibfnamefont {A.}~\bibnamefont
  {Mohorič}}, \bibinfo {author} {\bibfnamefont {G.}~\bibnamefont {Lahajnar}},\
  and\ \bibinfo {author} {\bibfnamefont {J.}~\bibnamefont {Stepišnik}},\
  }\href@noop {} {\bibfield  {journal} {\bibinfo  {journal} {Molecules}\
  }\textbf {\bibinfo {volume} {25}},\ \bibinfo {pages} {5813} (\bibinfo {year}
  {2020})}\BibitemShut {NoStop}%
\end{thebibliography}%

%\newpage

%\bibliography{mybib}


\end{document}